\documentclass{aa}  

\usepackage{graphicx}
\usepackage{txfonts}
\usepackage[hidelinks,colorlinks=true,linkcolor=blue,citecolor=blue]{hyperref}
\newcommand{\hi}{H{\sc i}}

\begin{document}

   \title{The BINGO Project VI:}

   \subtitle{\hi\ Halo Occupation Distribution and Mock Building}

   \author{Jiajun Zhang \inst{1},
Pablo Motta \inst{2}, 
Camila P. Novaes\inst{3},
Filipe B. Abdalla\inst{4}, 
Andre A. Costa\inst{5}, 
Bin Wang\inst{5,12},  
Zhenghao Zhu\inst{6},
Chenxi Shan\inst{6},
Haiguang Xu\inst{6,7},
Elcio Abdalla\inst{2}, 
Luciano Barosi\inst{8},
Francisco A. Brito\inst{8,13},
Amilcar Queiroz\inst{8},
Thyrso Villela\inst{3,11}, 
Carlos A. Wuensche\inst{3},
Elisa G. M. Ferreira\inst{2,9},
Karin S. F. Fornazier\inst{2},
Alessandro Marins\inst{2},
Larissa Santos\inst{5},
Marcelo Vargas dos Santos\inst{8},
Ricardo G. Landim\inst{10},
Vincenzo Liccardo\inst{3}
}

\institute{Center for Theoretical Physics of the Universe, Institute for Basic Science (IBS), Daejeon 34126, Korea\\
\email{liambx@ibs.re.kr}
\and
 Instituto de F\'isica, Universidade de S\~ao Paulo, C.P. 66.318, CEP 05315-970, S\~ao Paulo, Brazil 
\and
Instituto Nacional de Pesquisas Espaciais, Av. dos Astronautas 1758, Jardim da Granja, S\~ao Jos\'e dos Campos, SP, Brazil
\and 
Department of Physics and Astronomy, University College London, Gower Street, London WC1E 6BT, UK
\and
Center for Gravitation and Cosmology, College of Physical Science and Technology, Yangzhou University, Yangzhou 225009, China
\and
School of Physics and Astronomy, Shanghai Jiao Tong University, Shanghai 200240, China
\and
IFSA Collaborative Innovation Center, Shanghai Jiao Tong University, Shanghai 200240, China 
\and 
Unidade Acad\^emica de F\'isica, Universidade Federal de Campina Grande, R. Apr\'{\i}gio Veloso, 58429-900 - Bodocong\'o, Campina Grande - PB, Brazil    
\and
Max-Planck-Institut f{\"u}r Astrophysik, Karl-Schwarzschild Str. 1, 85741 Garching, Germany
\and
Technische Universit\"at M\"unchen, Physik-Department T70, James-Franck-Stra\text{$\beta$}e 1, 85748 Garching, Germany
\and
Instituto de F\'{i}sica, Universidade de Bras\'{i}lia, DF, Brazil
\and
School of Aeronautics and Astronautics, Shanghai Jiao Tong University, Shanghai 200240, China
\and
Departamento de F\'isica, Universidade Federal da Para\'iba, Caixa Postal 5008, 58051-970 Jo\~ao Pessoa, Para\'iba, Brazil
} 

   \date{Received xxxx, 2020; accepted xxxx, 2020}

 
    \titlerunning{The BINGO Project VI:: \hi\ Halo Occupation Distribution and Mock Building}
    \authorrunning{Jiajun Zhang et al.}

\abstract
{BINGO (\textbf{B}aryon Acoustic Oscillations from \textbf{I}ntegrated \textbf{N}eutral \textbf{G}as \textbf{O}bservations.) is a radio telescope designed to survey from 980 MHz to 1260 MHz, observe the neutral Hydrogen (\hi) 21-cm line and detect BAO (Baryon Acoustic Oscillation) signal with Intensity Mapping technique. Here we present our method to generate mock maps of the 21-cm Intensity Mapping signal covering the BINGO frequency range and related test results.}
{We would like to employ N-body simulations to generate mock 21-cm intensity map for BINGO and study what 21-cm intensity mapping observations can tell us about structure formation,\hi\ distribution and \hi\ mass-halo mass relation.}
{We have fitted an \hi\ mass-halo mass relation from ELUCID semi-analytical galaxy catalog and applied to Horizen Run 4 halo catalog to generate the 21-cm mock map, named as HOD. We have also applied the abundance matching method, matched the Horizen Run 4 galaxy catalog with \hi\ mass function measured from ALFALFA, to generate the 21-cm mock map, named as HAM}
{We studied the angular power spectrum of the mock maps and the corresponding pixel histogram. The comparison between two different mock map generation method (HOD and HAM) is presented. We provided the fitting formula of $\Omega_{\rm \hi\,}$, \hi\ bias and the lognormal fitting parameter of the maps, which can be used to generate similar maps. We discussed the possibility of measuring $\Omega_{\rm \hi\,}$ and \hi\ bias by comparing the angular power spectrum of the mock maps and the theoretical calculation. We also discussed the RSD (Redshift Space Distortion) effect, the nonlinear effect and the bin size effect in the mock map.}
{By comparing the angular power spectrum measured from two different kinds of mock maps and the theoretical calculation, we find that the theoretical calculation can only fit to the mock result at large scale. At small scales, neither the linear calculation nor the halofit nonlinear calculation can provide accurate fitting, which reflects our poor understanding of nonlinear distribution of \hi\ and its scale-dependent bias. We have found that the bias is highly sensitive to the method of populating \hi\ in halos, which also means we can put constraints on the \hi\ distribution in halos by observing 21-cm intensity mapping. We have also illustrated that only with thin frequency bins (such as 2 MHz), we can discriminate the effect of Finger-of-God. All of our investigations using mocks provide useful guide for our expectation of BINGO experiments and other 21-cm intensity mapping experiments.}

   {}


\maketitle


\section{Introduction}\label{sec:intro}
The spin-flip of the electrons in neutral hydrogen (\hi) emit or absorb photons in the wavelength of 21-cm and therefore indicate their location in the Universe. As a tracer of the underlying matter density field, 21-cm intensity mapping can reveal the tomographic Baryon Acoustic Oscillation signal in a cheap and fast fashion \citep{Wyithe2008MNRAS}. Therefore, several different experiments have been proposed to measure the 21-cm intensity map, such as CHIME \citep{chime}, Tianlai \citep{tianlai}, BINGO \citep{bingo}, FAST \citep{fast2,fast1,fast3}, SKA \citep{ska}, etc. By analyzing the large scale structure of the Universe using 21-cm intensity mapping, we can have better understanding about dark matter and dark energy \citep{statusreport}, which are crucial to understand the Universe. More detailed overview of BINGO project is described in the our companion papers I and II \citep{2020_project,BINGO.Instrument:2020}.

At low redshift \hi\ mostly contained inside galaxies and dark matter halos\citep{F18}. Since $z\sim 10$, the UV photons from the first generation stars started reionizing the Universe . At $z<5$, the Universe has been mostly ionized \citep{Becker2001AJ,Fan2006AJ,Fan2006ARA&A}. The \hi\ can be "protected" in high density regions\citep{Prochaska2005ApJ,Wolfe2005ARA&A,Zafar2013A&A}, due to sufficient column depth for self-shielding \citep{Pritchard2012}. The state-of-art hydrodynamic simulation IllustrisTNG \citep{F18} has shown that more than $95\%$ \hi\ gas is inside dark matter halo virial radius.

Therefore, it is valid to assume that \hi\ gas traces the distribution of galaxies and to use the distribution of galaxies and dark matter halos to estimate the distribution of \hi\ gas with high resolution N-body simulations. \citet{zhang2020parameter} has used the ELUCID simulation \citep{ELUCID3} together with a semi-analytical model \citep{Luoyu2016} to construct the \hi\ mock. They have carefully studied the Redshift Space Distortion (RSD) effect and shot noise for intensity mapping.

By observing the 21-cm emission line of \hi, we can determine the redshift of the emission source. Then, we can estimate the distance of the sources by applying Hubble's law. However, they are not the real distances of the sources. The redshift of the source is mainly contributed from the recessional velocity and the peculiar velocity. The estimation about the position of the source is distorted by the peculiar velocity. Such distortion is known as RSD effect. Since the peculiar velocity of the \hi\ gas is mainly affected by gravity from the clustering of matter, the RSD effect also contains rich information about the large scale structure. The RSD effect has been used to measure the growth factor of the Universe \citep{sdss4rsd}, test General Relativity \citep{Jullo2019A&A,fT2019PRD} and test different cosmological models\citep{Costa2017JCAP,An2019MNRAS,Yang2019PRD,Cheng2019}. The forecast of constraints from BINGO is described in our companion paper VII \citep{2020_forecast}.

The RSD effect can be classified into two parts, Kaiser effect \citep{kaiser1987} and Finger-of-God (FoG) effect \citep{FoG}. The moving of matter towards high density regions under gravity leads to the Kaiser effect. The Kaiser effect is dominant at large scales, which squeezes the distribution of galaxies in the line-of-sight direction. On the other hand, the random motion of matter inside high density regions introduces the FoG effect, which is dominant at small scales. It elongates the distribution of galaxies, making them look like fingers pointing at the observer. The RSD effect in 21-cm intensity mapping is well studied in \citet{Sarkar19}, and later improved in \citet{zhang2020parameter}.

There are various factors that contribute to the 21-cm intensity map, and the leading terms are the distribution of \hi\ gas in real space and the RSD effect \citep{HALL2013PRD}. The contributions from gravitational redshift, the integrated Sachs–Wolfe effect, gravitational lensing etc., can be neglected under current consideration. However, more detailed study of these effects might be necessary in the future experiments with higher sensitivity and resolution. Using the information of positions and velocity of halos and galaxies from simulations is sufficient to create a mock 21-cm intensity map for BINGO. 

In Sec. \ref{sec:overview}, we give the overview about the goals and methods of building mock map. We  introduce the \hi\ halo occupation distribution (HOD) model built from ELUCID simulation in Sec. \ref{sec:hod}. To validate the HOD model and the halo mass resolution, we test our method using ELUCID simulation in Sec. \ref{sec:tests}. In Sec. \ref{sec:mock}, we introduce the method of building a full-sky 21-cm intensity mock map. The measurements of the mock are presented in Sec. \ref{sec:result}. Finally we summarize the key points we have got in this study and discuss the future works in Sec. \ref{sec:summary}. This is the sixth of a series of  companion papers presenting the BINGO project. The companion paper I is the project paper \citep{2020_project}, paper II describes the instrument \citep{BINGO.Instrument:2020}, paper III the optics \citep{2020_optical_design}, paper IV simulations \citep{2020_sky_simulation}, paper V data analysis, correlations and component separation \citep{2020_component_separation}, and paper VII forecasts \citep{2020_forecast}.

\section{Overview}\label{sec:overview}

It is very popular in cosmological surveys to use simulated mocks to mimic real observations before the observations are taken. There are two major purposes for building the mock:
\begin{itemize}
    \item[1] Provide a mock data challenge, test the entire data analysis pipeline and see whether we can have the correct constraints of cosmological parameters.
    \item[2] Construct the covariance matrix, which is essential in understanding the source of error and uncertainty range of the parameter constraints.
\end{itemize}

For 21-cm intensity mapping, the data product of the mock is the brightness temperature distribution in 2D spherical surface at different redshifts. Such data product consists of the following four components:
\begin{itemize}
    \item[1] signal, the 21-cm emission from extra-galactic sources, which mainly comes from \hi\ gas in the galaxies and dark matter halos.
    \item[2] foreground, coming from radio wave emission in the Milky Way and the extra-galactic sources, which mainly consists of synchrotron radiation and free-free emission.
    \item[3] noise, including thermal noise, shot noise and other sources of noise.
    \item[4] mask, cover part of the sky for removing too strong foreground, usually around very bright stars or directly toward the Milky Way, or to account for non observed regions.
\end{itemize}

In this paper, we will only discuss about the 21-cm signal. The other three components are discussed in our companion paper IV and V \citep{2020_sky_simulation,2020_component_separation}. In order to generate the mock 21-cm intensity map, the key part is determining the distribution of \hi. There are 6 ways to generate the \hi\ gas distribution:
\begin{itemize}
    \item[1] Hydro: Directly read out the gas particle properties from hydrodynamic simulations, such as IllustrisTNG \citep{F18}.
    \item[2] SAM: Use the \hi\ gas mass information from semi-analytical galaxy catalog based on N-body simulations, such as ELUCID \citep{zhang2020parameter}.
    \item[3] N-body: Use empirical \hi\ mass-halo mass relation to populate \hi\ gas in dark matter halos, such as HIR4 \citep{asorey2020hir4}.
    \item[4] Fast Halo: Again, relate \hi\ mass with halo mass using empirical relations, but the dark matter halo catalog is generated by fast simulation method, such as {\tt COLA-HALO} \citep{koda2016fast}.
    \item[5] Model: Populate \hi\ mass on lognormal dark matter density distribution \citep{alonso2014fast,Xavier:2016} or even Gaussian realization\citep{asorey2020hir4}.
    \item[6] Poisson Halo: Use empirical \hi\ mass-halo mass relation to populate \hi\ gas in dark matter halos, but the sub-resolution dark matter halos are generated by Poisson sampling on dark matter density fields measured in N-body simulations\citep{seehars2016JCAP}.
\end{itemize}

Clearly, Hydro is the most expensive and accurate method. However, Hydro is too expensive to construct a full sky mock. Even the state-of-art hydrodynamic simulation cannot fulfill our requirement. Semi-analytical galaxy catalog is built from merger-tree traced by N-body simulations, which requires subhalo identification. SAM method is less accurate than hydrodynamic simulation, but more efficient. Populating \hi\ mass in dark matter halos identified from N-body simulation or fast simulations is even more efficient, but less accurate. Using dark matter halos Poisson sampled from dark matter density field can largely reduce the requirement of simulation resolution, which is very useful for large scale application. The lognormal (Gaussian) density generation method is the most efficient among these six methods, but it is the least accurate one. We need to find the balance between accuracy and efficiency for different purpose of mock building. We also need to notice that, in order to construct a full-sky light cone from $z=0.12$ to $z=0.45$ without using periodic boundary, the box size of such simulation will be larger than $2.4 \text{Gpc}/h$. Currently, we do not have such large scale hydrodynamic simulation. Therefore, in this paper, we will introduce our empirical relations from IllustrisTNG and ELUCID semi-analytical model, and apply that onto Horizon Run 4 simulation \citep{kim2015horizon} light cone to generate the mock map. This method can also be applied on fast halo generation catalogs for calculating covariance matrix in the future.
\section{\hi\ Halo Occupation Distribution}\label{sec:hod}
\subsection{ELUCID SAM catalog}\label{subsec:elucid}
\begin{figure*}
    \centering
    \includegraphics[width=0.35\textwidth]{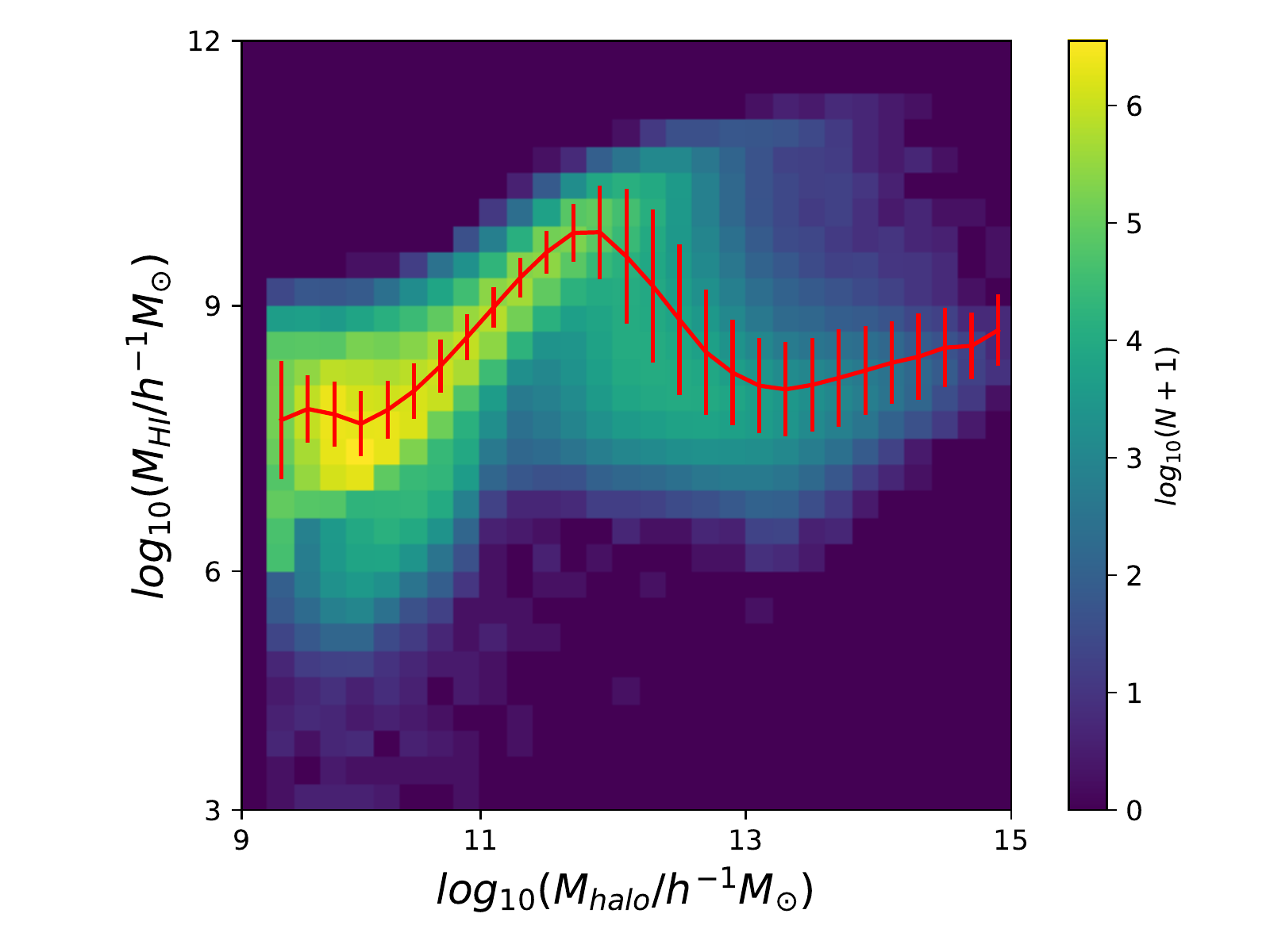}
    \includegraphics[width=0.35\textwidth]{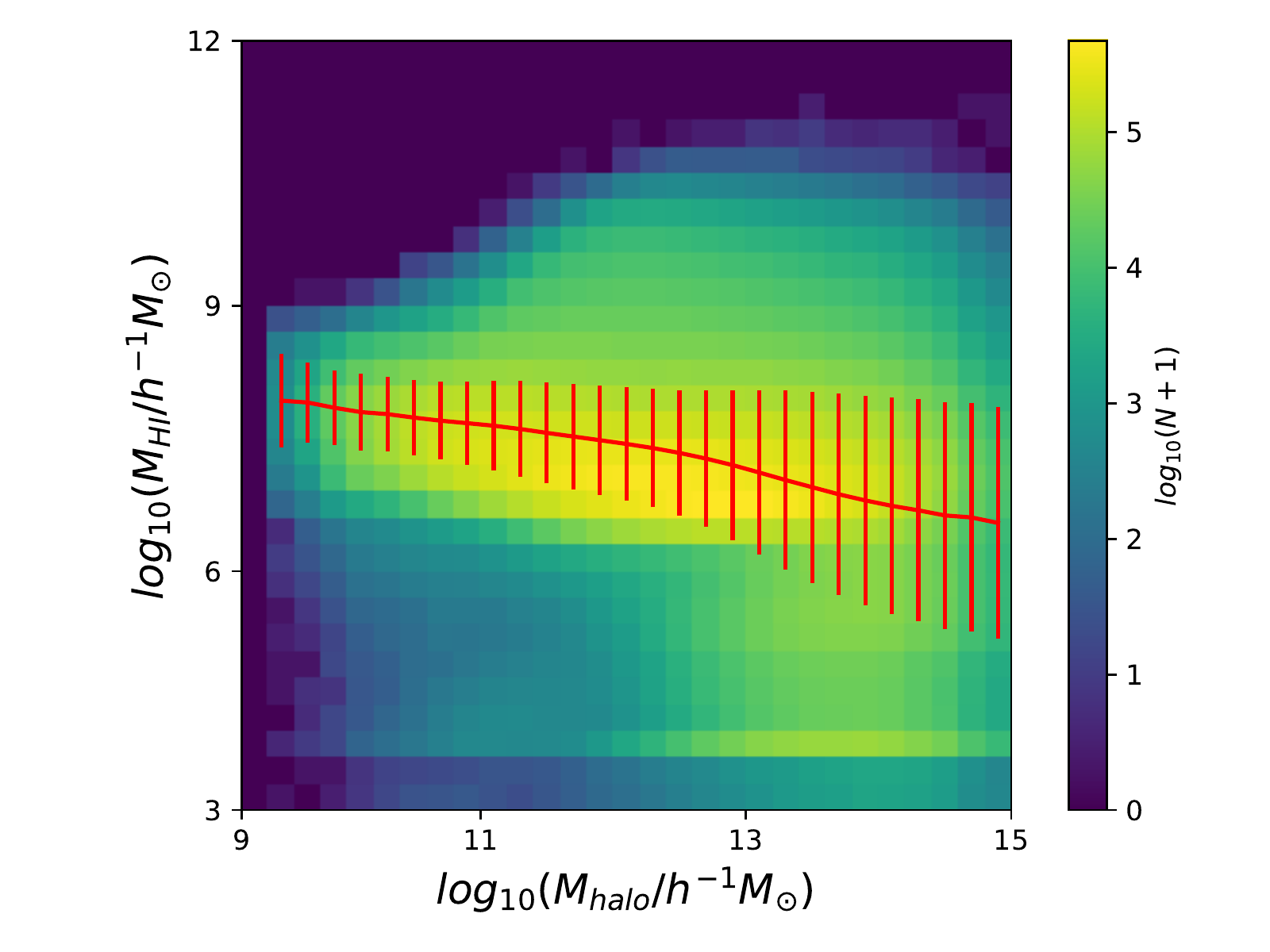}
    \includegraphics[width=0.265\textwidth]{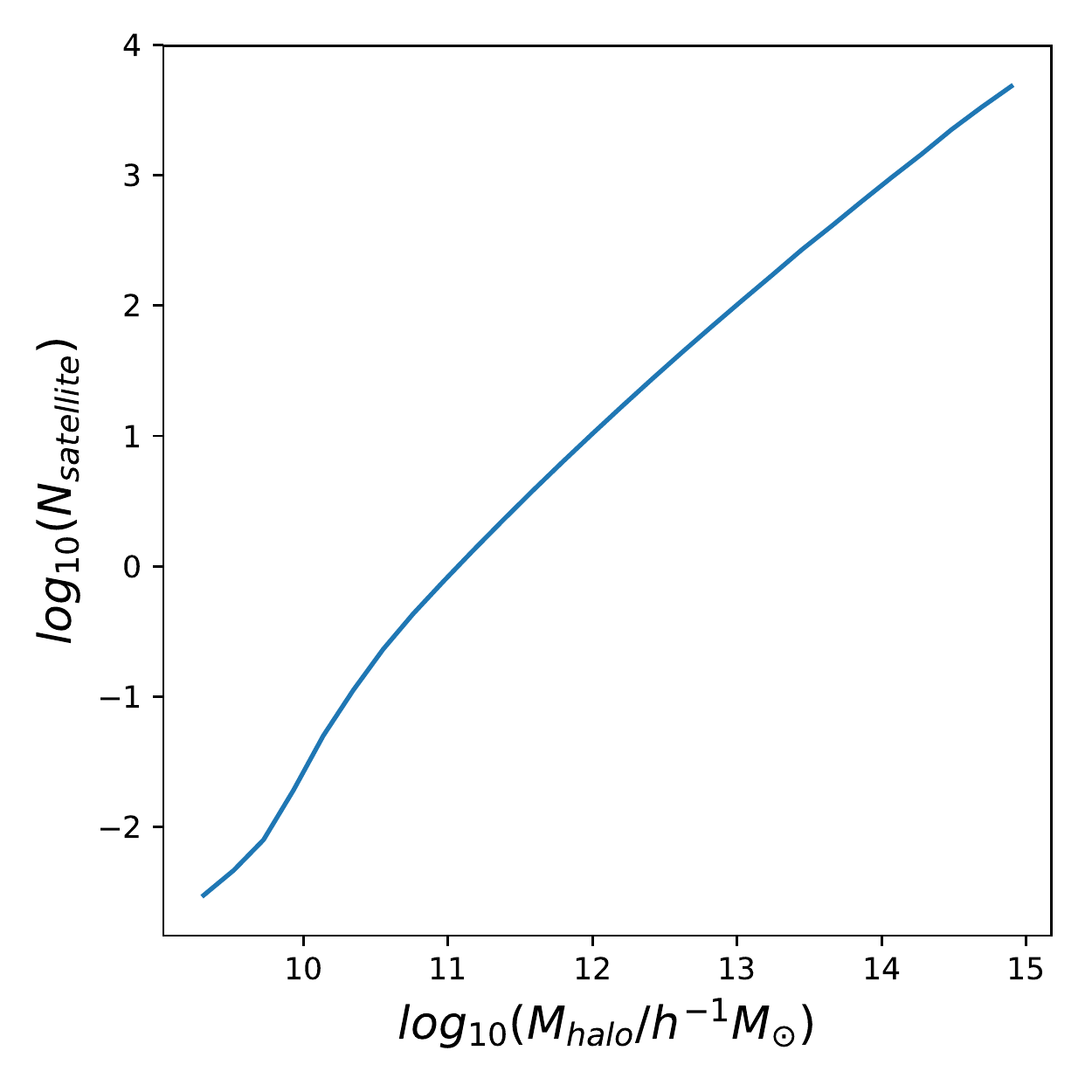}
    \caption{We show the number of galaxies as a function of their host halo mass and \hi\ mass in the ELUCID SAM catalog. The number in logarithmic scale is represented by different colors. We show the result of central (satellite) galaxies on the left (middle) panel. The mean value and standard deviation are shown together with the color map to illustrate the trend more clearly. The average number of satellite galaxies in each halo as a function of mass is shown on the right panel.}
    \label{fig:hod1}
\end{figure*}
We use the semi-analytical galaxy catalog from ELUCID N-body simulation \citep{ELUCID3,Luoyu2016}. The ELUCID simulation is an N-body simulation with $3072^3$ particles in a periodic cubic box of $500h^{-1}$Mpc on a side. WMAP5 cosmology \citep{wmap5} was assumed in the ELUCID simulation. By following the the merger tree of dark matter halos in the simulations, \citet{Luoyu2016} constructed the galaxy catalog using their semi-analytical model, which considered the physical process like gas cooling, star formation, supernova feedback, etc. In this work for studying the \hi\ gas halo occupation distribution, we use the position, velocity, \hi\ mass and dark matter halo mass of the galaxies from the galaxy catalog, the lower limit of the dark matter halo mass is about $1.85\times10^9 M_{\odot}/h$. We assume that all the \hi\ mass in the Universe is inside galaxies and their hosting halos, concentrated in the center of the galaxies. As described in \citet{F18}, from IllustrisTNG simulation \citep{tng1,tng2,tng3,tng4,tng5} results, it was shown that at $z<3.0$, more than $90\%$ of the \hi\ gas is inside the galaxies and more than $95\%$ of the \hi\ mass is inside the halos. Therefore, this assumption is reasonable.

In this galaxy catalog, we also have the information about whether the galaxy is the central galaxy of the halo or a satellite galaxy. This allows us to further separate the galaxies into central galaxies and satellite galaxies for further discussion. Although IllustrisTNG is the state-of-art hydrodynamic simulation, its box size is still quite small (at most $205\,\mathrm{Mpc}/h$ for IllustrisTNG-300 simulation and smaller for the other IllustrisTNG simulations). Therefore, a larger box size simulation like ELUCID with high particle resolution can provide better statistics about the \hi\ halo occupation distribution in very massive clusters. The 21-cm intensity mapping using a semi-analytical model galaxy catalog from MDPL2 simulation has been considered \citep{cunnington2020multipole}. However the MDPL2 simulation can only resolve halos with mass $>3\times10^{10} M_{\odot}/h$ \citep{klypin2016multidark}, one order of magnitude larger than the ELUCID simulation. Without sufficient resolution, it will be hard to tell the \hi\ distribution in low mass halos like $10^9 M_{\odot}/h$. The ELUCID SAM (semi-analytical model) catalog is a good choice for studying the \hi\ halo occupation distribution in both the high mass end and low mass end. We will use this catalog to study the \hi\ distribution in halos throughout this section.  
\subsection{Central and Satellite Galaxies}\label{subsec:galaxy}

In general, gas cannot efficiently cool and form galaxies in both too massive dark matter halos or too light dark matter halos. However, due to different formation history, central galaxies and satellite galaxies can host very different fraction of \hi\ gas. \citet{padmanabhan2017himass} provided a halo model and an \hi\ mass-halo mass relation, but the subhalos inside main halos was not taken into account separately. An \hi\ mass-halo mass relation was provided by \citet{Guo2017ApJ}, which was fitted to $\sim16000$ galaxies in the range of $0.0025< z< 0.05$ from ALFALFA. The importance of satellite galaxies has been illustrated in an updated halo model by \citet{Paul2018MNRAS}. By studying the different \hi\ gas distribution in satellite galaxies and central galaxies using the SAM galaxy catalog at different redshift, we hope to provide a different \hi\ mass-halo mass relation, which is more suitable for our redshift range. We show the number of galaxies as a function of their \hi\ mass and host halo mass in Fig. \ref{fig:hod1}, measured from ELUCID SAM catalog  at $z=0$. We can see that for central galaxies, it is more abundant with less massive host halo and less \hi\ mass. It is easy to understand because less massive halos are more abundant, and \hi\ mass 
positively correlates to the halo mass if the halo mass is not more massive than $\sim10^{12} M_{\odot}/h$.
\begin{figure}
    \centering
    \includegraphics[width=0.45\textwidth]{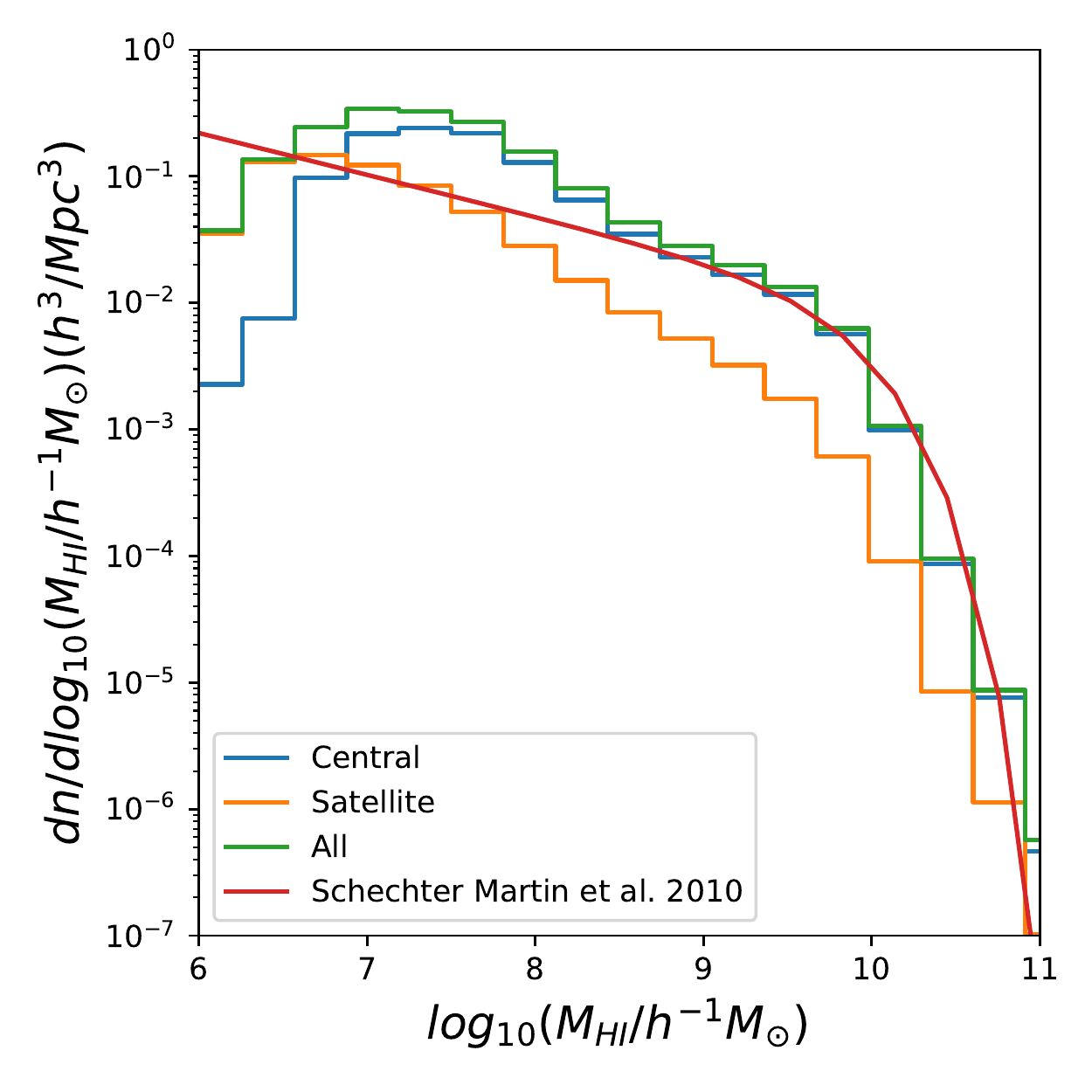}
    \caption{The \hi\ mass function measured from the ELUCID SAM catalog at $z=0$ is shown. The result of central (satellte) galaxy is shown in blue (orange), and the total result is shown in green. The red curve shows the Schechter function provided by \citet{Martin2010ALFALFA}, which represents the fitting function of ALFALFA observation. Overall, the \hi\ mass function between the SAM and observation is consistent.}
    \label{fig:hod2}
\end{figure}

Together with the number of galaxies represented by different color, we have also shown the mean value and the standard deviation in Fig.\ref{fig:hod1}. We can see the mean \hi\ mass of the central galaxies clearly has a peak around host halo mass $M_{\rm halo}=10^{11.6} M_{\odot}/h$. Below that, \hi\ mass positively correlates to the halo mass and beyond that peak, \hi\ mass drops with larger halo mass until $M_{\rm halo}\sim10^{13} M_{\odot}/h$. The mean \hi\ mass of satellite galaxies depends weakly on the halo mass. With larger halo mass, the number of satellite galaxies rises up quickly, and the standard deviation of their \hi\ mass also increase. Therefore, the total \hi\ mass in a dark matter halo consists of the contribution of the central galaxies and satellite galaxies differently for different halo masses. 

The \hi\ mass function measured from the ELUCID SAM catalog at $z=0$ is shown in Fig.\ref{fig:hod2}. We have shown the mass function of satellite galaxies and central galaxies separately, as well as the total \hi\ mass function. The Schechter function given in \citet{Martin2010ALFALFA} is shown in red curve for comparison. The SAM \hi\ mass function can well fit the Schechter function at $M_{HI}>10^{8}h^{-1}M_{\odot}$ but deviate from it at low mass end. Overall, the SAM catalog still provides a reasonably similar mass function to that fitted from ALFALFA observation.
\subsection{\hi\ Mass Halo Mass relation}\label{subsec:hihalo}
\begin{figure}
    \centering
    \includegraphics[width=0.45\textwidth]{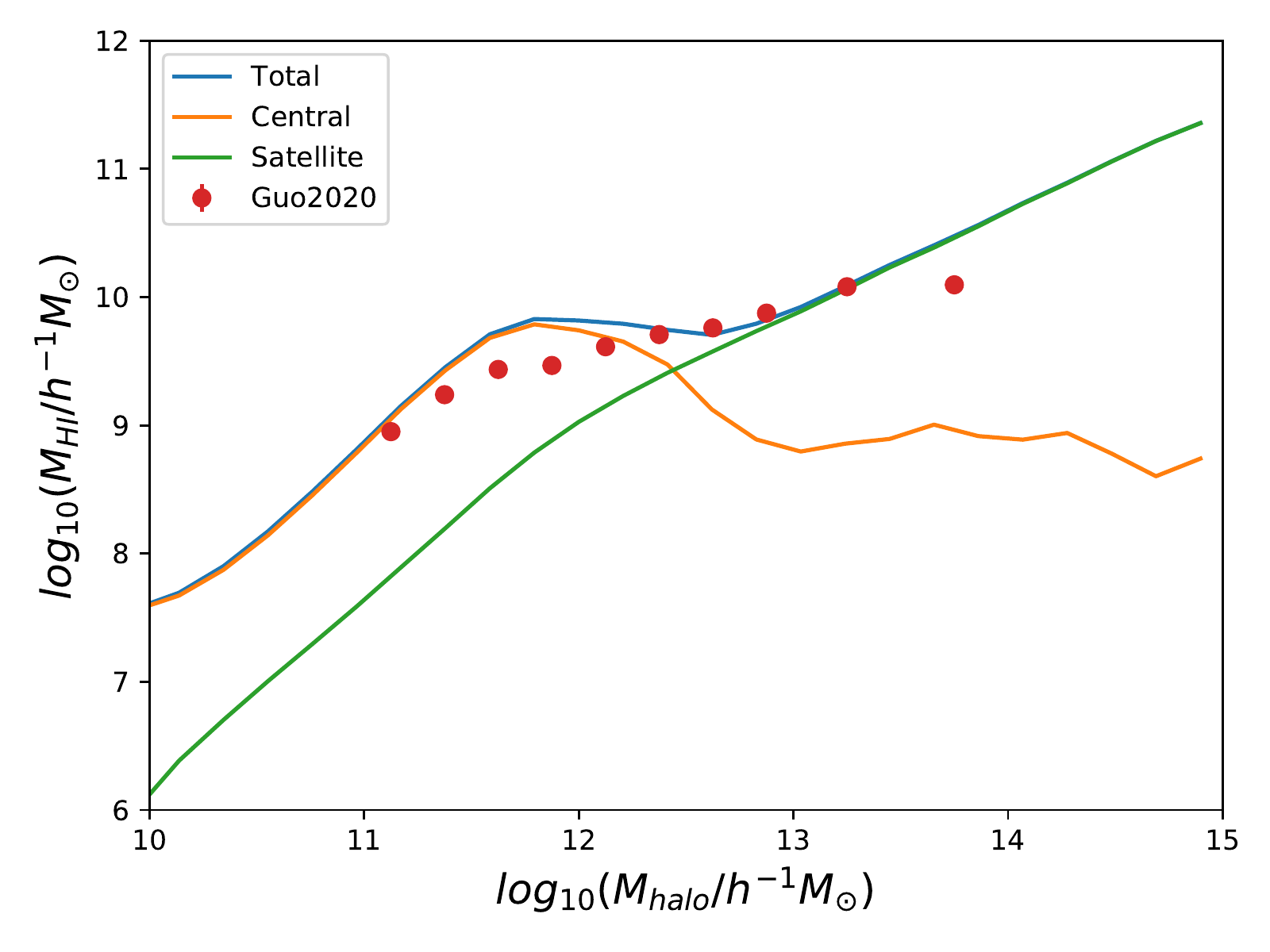}
    \includegraphics[width=0.45\textwidth]{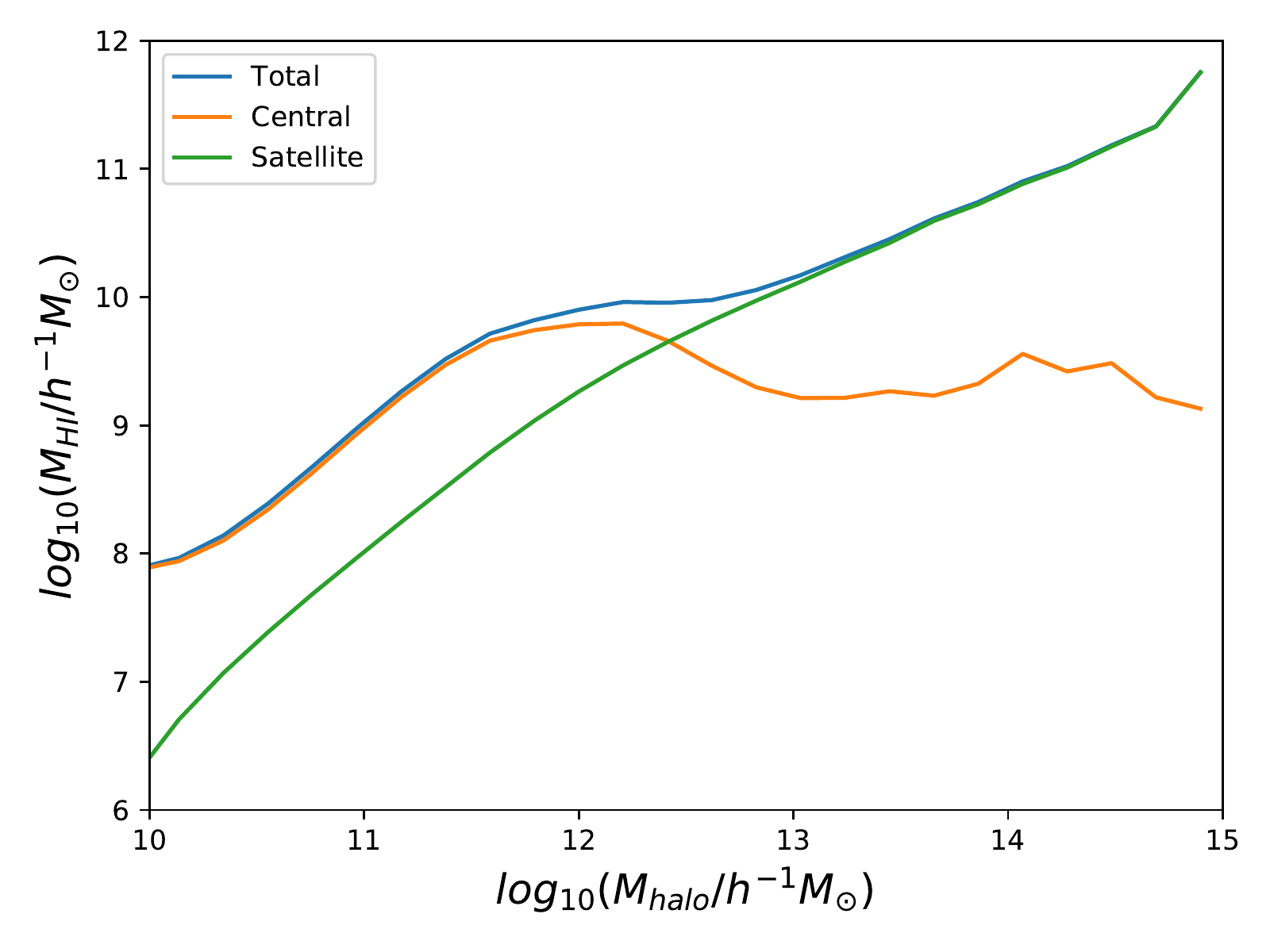}
    \caption{In the upper(lower) panel, we show the \hi\ mass-halo mass relation at $z=0 (0.66)$. The total \hi\ mass is shown in blue curve, the contribution from central galaxies is shown in orange and the contribution from satellite galaxies is shown in green. No matter z = 0 or 0.66, the central galaxy always contributes the most \hi\ mass in low mass halos and satellite galaxies contribute the most \hi\ mass in high mass halos. We have also shown the observed \hi\ mass-halo mass relation in red dots at $z=0$ \citep{guo2020hihalo} for comparison.}
    \label{fig:hihalo1}
\end{figure}

For generating mock 21-cm map, the data product from N-body simulations or fast simulations is usually dark matter halo catalog. The most popular and simplest information in the halo catalog are the position, the velocity and the mass of the halo. Therefore, it is most useful if we can find an \hi\ mass-halo mass relation, providing \hi\ mass as a function of halo mass. We can see in Fig. \ref{fig:hihalo1} that, for low mass halos, the major contribution of \hi\ mass is from the central galaxies, and for high mass halos, the major contribution comes from satellite galaxies. Although in massive halos, there are more \hi\ mass in central galaxies than in every individual satellite galaxies on average, the number of satellite galaxies is so huge that they still contribute the most \hi\ mass in massive halos. It is very important to consider the satellite galaxy contribution of \hi\ mass in massive halos, which is ignored in HIR4 mock \citep{asorey2020hir4}. We have also compared the \hi\ mass-halo mass relation to the observational results \citep{guo2020hihalo} at $z=0$, the results from ELUCID galaxy catalog fits relatively well to the observation.
\begin{figure}
    \centering
    \includegraphics[width=0.45\textwidth]{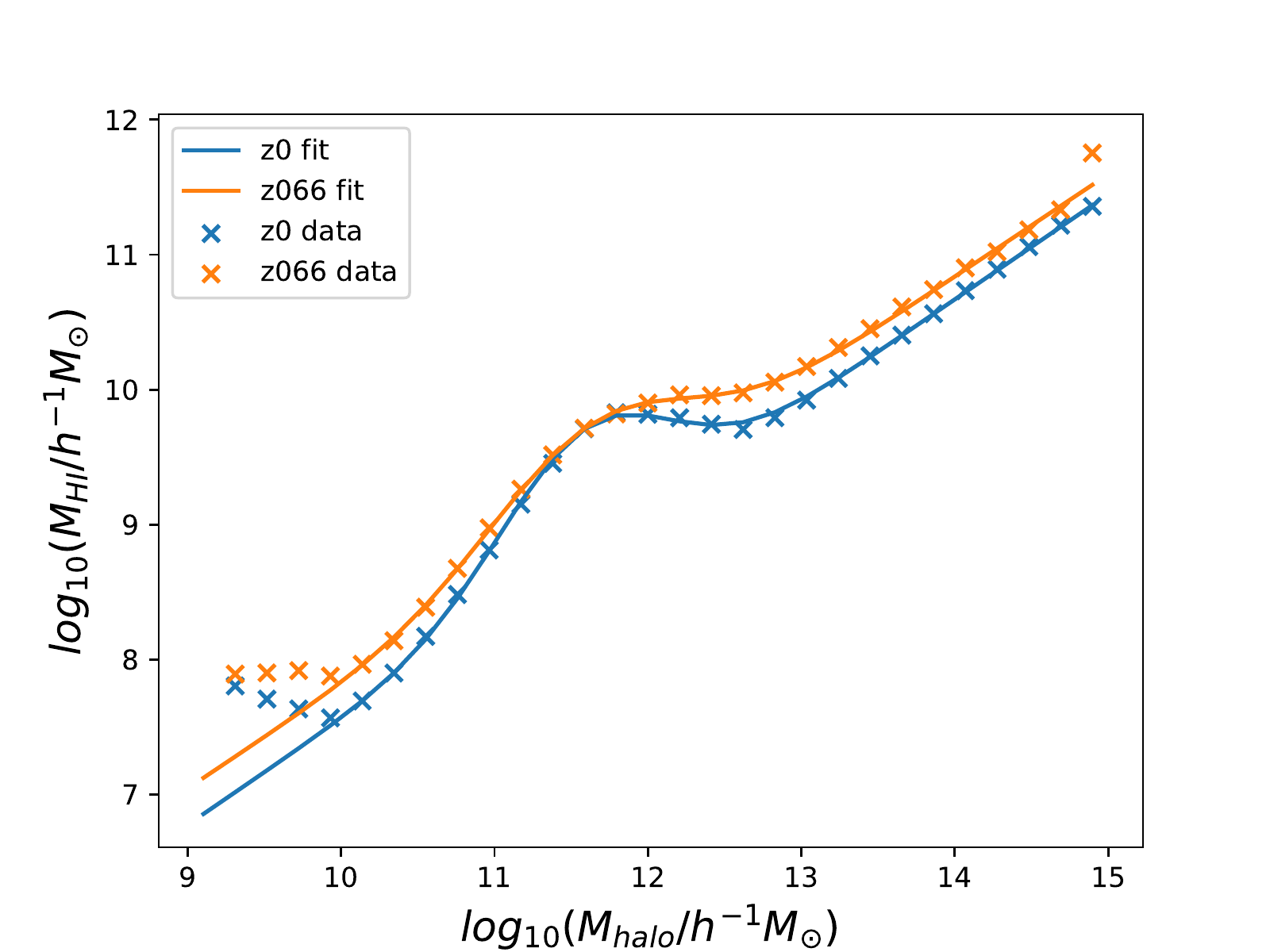}
    \caption{We fit the measured \hi\ mass-halo mass relation from ELUCID SAM catalog with the analytical formula in solid lines. We have chosen the best fit curve in the mass range $10^{10} M_{\odot}/h<M_{halo}<10^{15} M_{\odot}/h$.}
    \label{fig:hihalo2}
\end{figure}

We first fitted the measured \hi\ mass-halo mass relation from ELUCID SAM catalog using a linear function plus a Gaussian function with free parameter $a$, $b$, $c$, $d$ and $f$, then we get the parameters as functions of redshift $z$, by assuming that the parameters are linearly related with redshift in $0<z<0.66$. Then we find the following analytical formula, which can be easily applied to halo catalogs with the information of halo mass,
\begin{equation}\label{eq:hihalofit}
\begin{aligned}
    &\log_{10}(m_{\rm \hi\,}) = a\log_{10}(m)+b+c e^{-(\log_{10}(m)-d)^{2}/f^2}\,,\\
    &a = 0.78 - 0.030z\,,\\
    &b = -0.23 + 0.68z\,,\\
    &c = 0.92 - 0.32z\,,\\
    &d = 11 + 0.038z\,,\\
    &f = 0.79 + 0.18z\,,
\end{aligned}
\end{equation}
where $z$ is the redshift, $m_{\rm \hi\,}$ is the \hi\ mass in unit of $M_{\odot}/h$, $m$ is the halo mass in unit of $M_{\odot}/h$. The halo mass definition is based on FoF (Friend-of-Friend) halo finding algorithm, the linking length is $0.2$ times the mean particle separation distance. The fitting result is shown in Fig. \ref{fig:hihalo2} Notice that the \hi\ mass-halo mass relation is very similar between $z=0$ and $z=0.66$, therefore, the assumption that the parameters are linearly related with redshift in $0<z<0.66$ is reasonable.
\section{Tests with ELUCID}\label{sec:tests}
\subsection{The Halo Mass Threshold}\label{subsec:threshold}
\begin{figure*}
    \centering
    \includegraphics[width=0.45\textwidth]{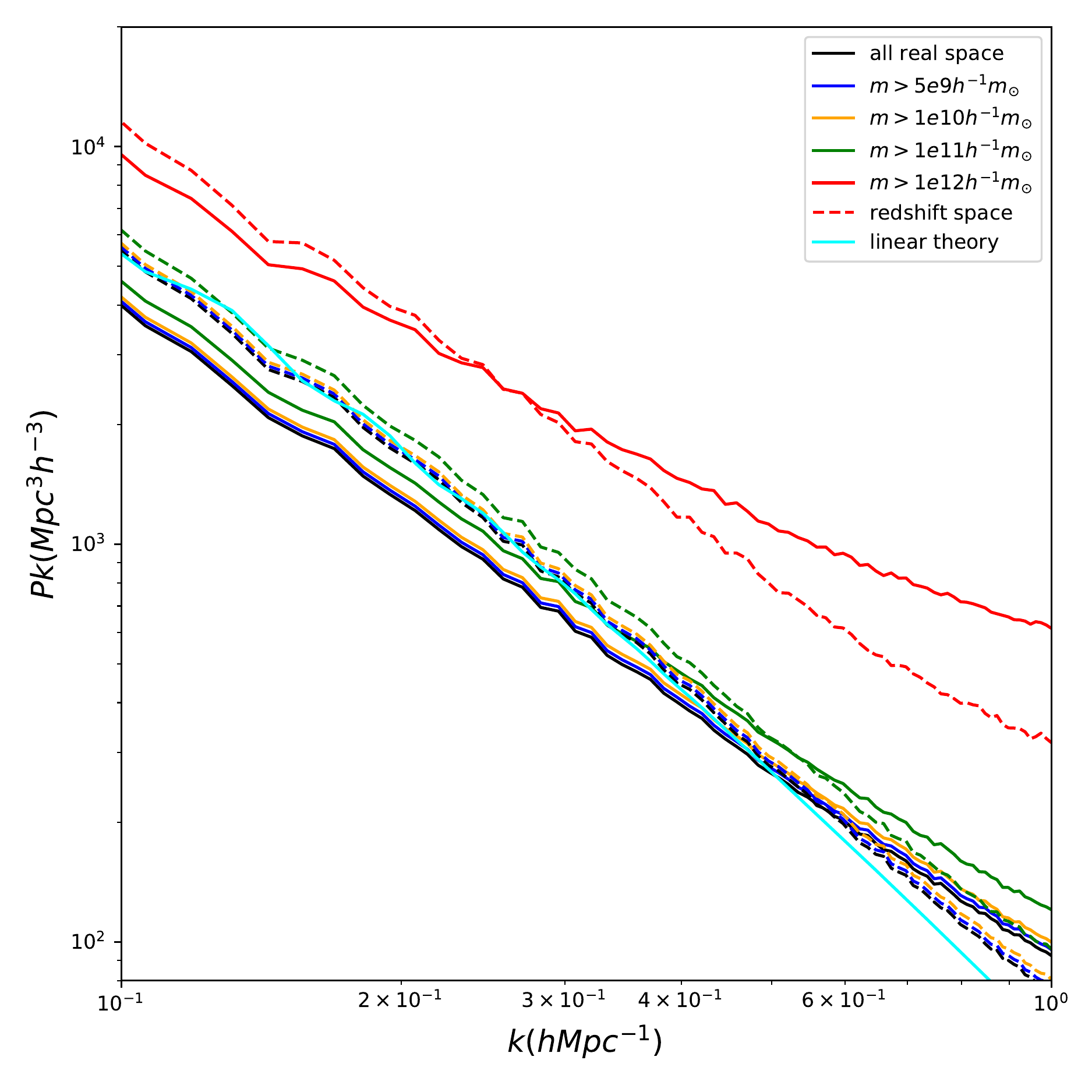}
    \includegraphics[width=0.45\textwidth]{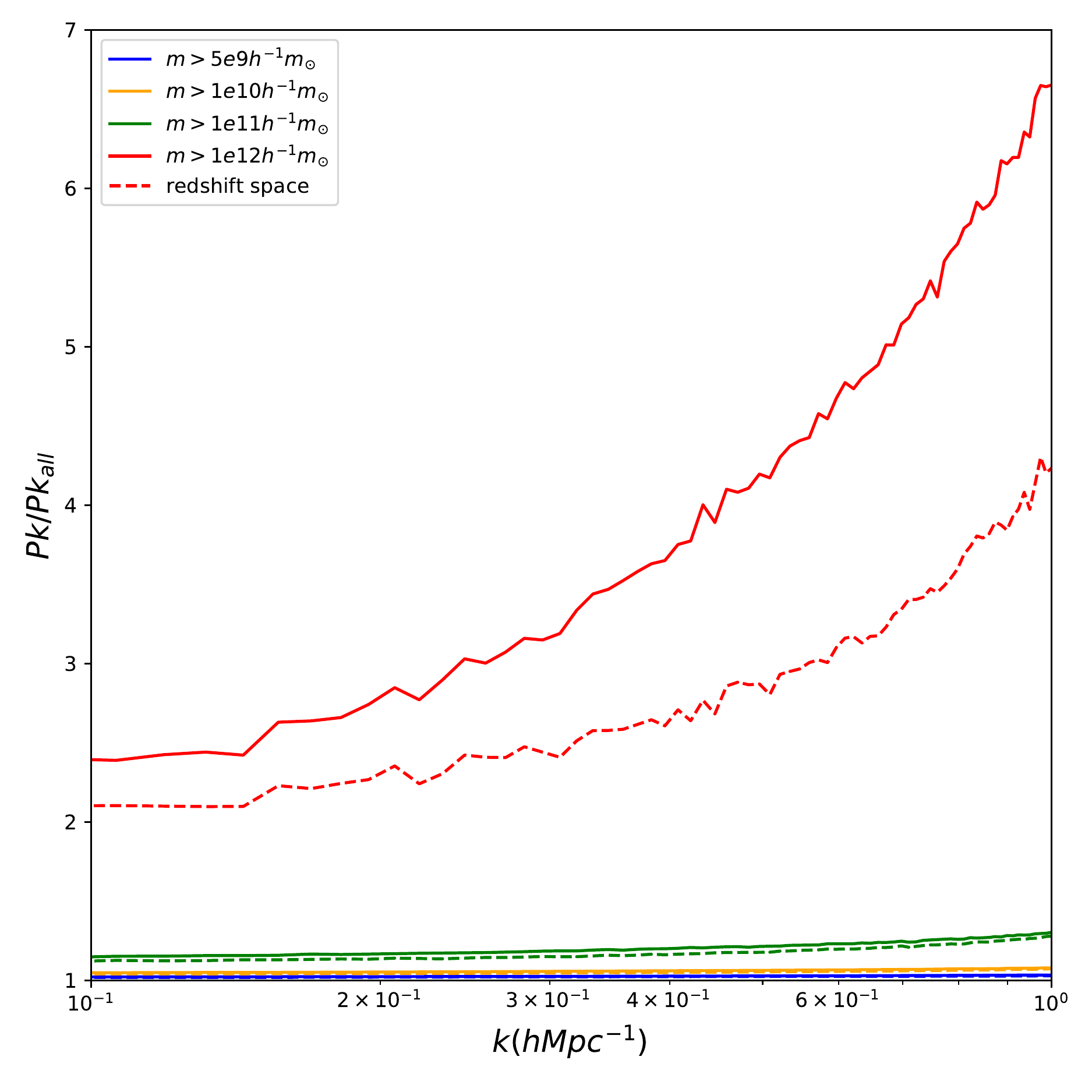}
    \caption{On the left, we show the \hi\ power spectrum and on the right, we show the ratio of power spectrum between the sample with halo mass threshold and the full sample. The solid curves show the results in real space and the dashed curves show the results in redshift space. Different colors denote different samples with different threshold. We have also included the linear matter power spectrum for reference in cyan color on the left.}
    \label{fig:threshold}
\end{figure*}

When building the mock \hi\ distribution, the mass resolution and box size of the halo catalog we use is usually limited. We need to know what is the effect of limited mass resolution on the \hi\ power spectrum. Such effect can be quantified by the bias of \hi\ gas in the mass limited sample. The \hi\ gas traces the matter distribution in a biased way, we usually use the bias parameter $b_{\rm \hi\,}$ to describe the difference between \hi\ distribution and the matter distribution. Similarly, the dark matter halos also trace matter distribution in a biased way, and the halo bias is $b$. It has been addressed in HIR4 \citep{asorey2020hir4} that the bias of \hi\ power spectrum can be calculated as the weighted average of halo bias,
\begin{equation}
    b_{\rm \hi\,}(z)=\dfrac{\int dm\, n(m,z)m_{\rm \hi\,}(m,z)b(m,z)}{\int dm\, n(m,z)m_{\rm \hi\,}(m,z)}\,,
\end{equation}
where $b(m,z)$ represents the halo bias, $m_{\rm \hi\,}$ is the \hi\ mass and $m$ is the halo mass. The halo bias can be calculated by empirical formula calibrated by N-body simulations \citep{tinker2010bias}. Notice that the halo bias used here is scale independent, which means the resulted \hi\ bias model is also scale independent. In other words, if the \hi\ bias is actually scale dependent, our bias model cannot provide a proper description. 

With the high resolution ELUCID simulation, we can shed light on whether a scale independent bias model is a proper description or not. We selected the galaxies according to their host halo mass with the mass threshold $m>5\times10^{9}M_{\odot}/h,1\times10^{10}M_{\odot}/h,1\times10^{11}M_{\odot}/h,1\times10^{12}M_{\odot}/h$. In Fig. \ref{fig:threshold}, we show the \hi\ power spectrum of the selected samples and the full sample on the left panel, while on the right panel we show the ratio of \hi\ power spectrum of the selected samples and the full sample. The linear matter power spectrum calculated using \citet{EH1998ApJ} is shown for reference in cyan color on the left panel. Considering the RSD effect, the positions of the galaxies ($\vec{x}$) are mapped to redshift space ($\vec{s}$) using the plane-parallel approximation
\begin{equation}
    \vec{s}=\vec{x}+\dfrac{1+z}{H(z)}\vec{v}_{\rm los}\,,
\end{equation}
where $\vec{v}_{los}$ is the peculiar velocity of the galaxies along the line-of-sight. The redshift space results are shown in dashed lines in Fig. \ref{fig:threshold}. Notice that we have deducted the shot noise power for the illustration and calculation of the power spectrum ratio. Only when the ratio of power spectrum is close to a constant, we can safely use the scale independent halo bias to calculate the \hi\ bias. Considering the scale of 1 degree in the sky, which corresponds to $k\sim0.5h \text{Mpc}^{-1}$ at $z\sim0.12$, the power spectrum ratio of $m>10^{11}M_{\odot}/h$ from $k\sim0$ to $k\sim0.5h\text{Mpc}^{-1}$ is close to a constant, but in the $m>10^{12}M_{\odot}/h$ case, the ratio cannot be well represented by a constant. 

Therefore, we conclude that the halo mass resolution for making a mock 21-cm intensity map requires at least $\sim10^{11}M_{\odot}/h$. The finer the mass resolution, the better.
\subsection{Repopulate \hi\ Mass}\label{subsec:repopulate}
\begin{figure*}
    \centering
    \includegraphics[width=0.45\textwidth]{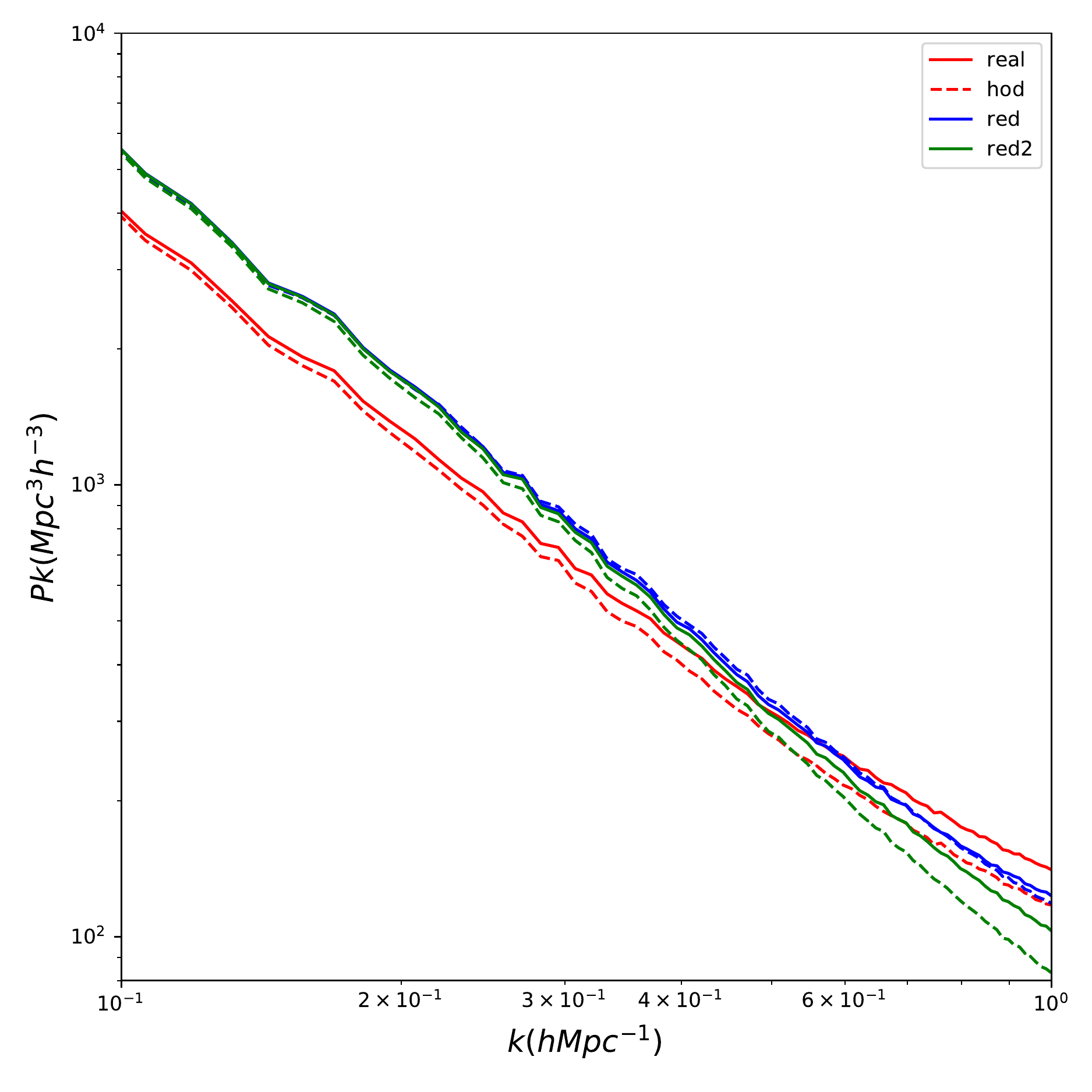}
    \includegraphics[width=0.45\textwidth]{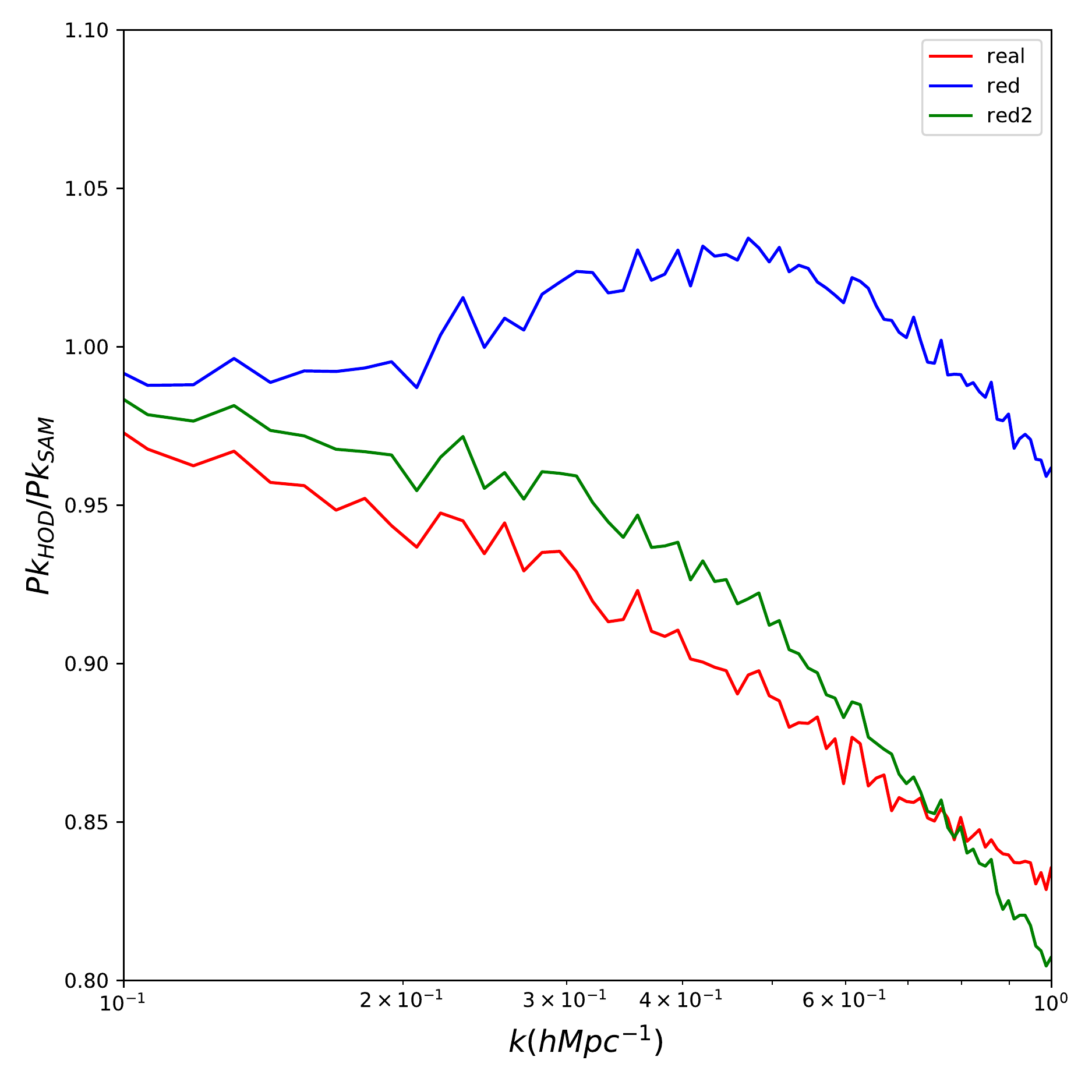}
    \caption{On the left, we show the \hi\ power spectrum, the solid lines represent the power spectrum measured from the SAM galaxy catalog, the dashed lines represent the power spectrum measured from the HOD model. "real" means real space point mass power spectrum, "red" means redshift space point mass power spectrum and "red2" means redshift space inside-galaxy \hi\ distribution modelled power spectrum \citep{zhang2020parameter}. On the right, we show the power spectrum ratio between the results of HOD model and SAM galaxy catalog.}
    \label{fig:hodpk}
\end{figure*}

We use the \hi\ mass-halo mass relation introduced in Sec. \ref{subsec:hihalo} to repopulate the main halos in ELUCID simulation at $z=0$ with \hi\ gas, in order to test the performance of our \hi\ HOD model. We have generated three kinds of \hi\ maps and calculated their power spectrum, 
\begin{itemize}
    \item[1] Use the galaxy (halo) positions in real space, treat the galaxies (halos) as point mass and calculate the \hi\ distribution in the SAM (HOD) catalog,
    \item[2] Use the galaxy (halo) positions in redshift space, treat the galaxies (halos) as point mass and calculate the \hi\ distribution in the SAM (HOD) catalog,
    \item[3] Use the galaxy (halo) positions in redshift space, treat the galaxies (halos) as extended mass due to \hi\ velocity dispersion inside galaxies (halos) and calculate the \hi\ distribution in the SAM (HOD) catalog.
\end{itemize}

We modelled the \hi\ velocity dispersion inside galaxies at $z=0$ as,
\begin{equation}\label{eq:sigmav0}
    \sigma_{v} = \dfrac{31 \text{km/s}}{\sqrt{3}}\left(\dfrac{m}{10^{10}M_{\odot}/h}\right)^{0.35},
\end{equation}
which is also used in \citet{zhang2020parameter}, derived from IllustrisTNG simulation \citep{F18}. Usually, we do not consider the \hi\ velocity dispersion inside galaxies (halos) when making the mock 21-cm intensity map using N-body simulations. \citet{zhang2020parameter} has pointed out that with the proper consideration of such \hi\ velocity dispersion, we can have a more realistic description of \hi\ mass distribution in redshift space. Therefore, we have made these three kinds of \hi\ maps for testing. The results are shown in Fig. \ref{fig:hodpk}. The power spectrum is shown on the left panel and the ratio between HOD model and SAM catalog is shown on the right panel. On the left panel, the dashed lines represent the \hi\ power spectrum measured from HOD model and the solid lines show the results of the original SAM catalog. We can see that overall, the HOD model can recover the \hi\ power spectrum in the SAM catalog within $10\%$ ($20\%$) accuracy up to $k=0.5h\text{Mpc}^{-1}$ ($k=0.8h \text{Mpc}^{-1}$). Therefore, we have validated the method of the HOD model to populate dark matter halos with \hi\ mass.
\section{Mock Building with HR4}\label{sec:mock}
\subsection{Horizon Run 4 Catalog}\label{subsec:hr4}

Horizon Run 4 (HR4 in short) is a cosmological N-body simulation with cubic box size of $3150\text{Mpc}/h$ and $6300^3$ number of particles \citep{kim2015horizon}. HR4 simulation adopted WMAP5 $\Lambda$CDM cosmology, where $\Omega_m=0.26,\Omega_\Lambda=0.74,h=0.72,\sigma_8=0.79,n_s=0.96$ \citep{wmap5}. The halos are identified by Friend-of-Friend (FoF) algorithm with linking length equal to 0.2 times the particle mean distance in HR4 simulation, at least 30 particles are required for identifying a halo, which means the minimum halo mass is $2.7\times10^{11}M_{\odot}/h$. 

We use the halo light cone catalog and the mock galaxy light cone catalog to generate our mock 21-cm intensity map. Usually, in order to generate a full sky light cone catalog, we will use the periodic boundary conditions of the simulations to pile up simulation boxes together to get a deep light cone. This technique will introduce some duplicate structures. However, the HR4 simulation box size is so large that there is no periodic structures in $z<0.6$ in the full sky light cone catalog. 

The redshift range of BINGO is $0.127<z<0.449$, therefore, there is no duplicate structures in our mock map. In summary, there are four steps in making the mock map,
\begin{itemize}
    \item[1] Populate \hi\ gas in the galaxy (halo) catalog and calculate the \hi\ density distribution $\rho_{\rm \hi\,}$
    \item[2] Calculate the brightness temperature 
    \begin{equation}\label{eq:Tb}
        T_b=189h\dfrac{H_0 (1+z)^2}{H(z)}\dfrac{\rho_{\rm \hi\,}}{\rho_{c}}\text{mK}\,,
    \end{equation} where $\rho_c$ is the critical density of the Universe \citep{F18},
    \item[3] Re-scale the map according to required $\Omega_{\rm \hi\,}$,
    \item[4] Smooth the brightness temperature map with a Gaussian kernel FWHM = 40 arcmin, which is due to the beam of BINGO. 
\end{itemize}
\subsection{Abundance Matching}\label{subsec:am}
\begin{figure}
    \centering
    \includegraphics[width=0.45\textwidth]{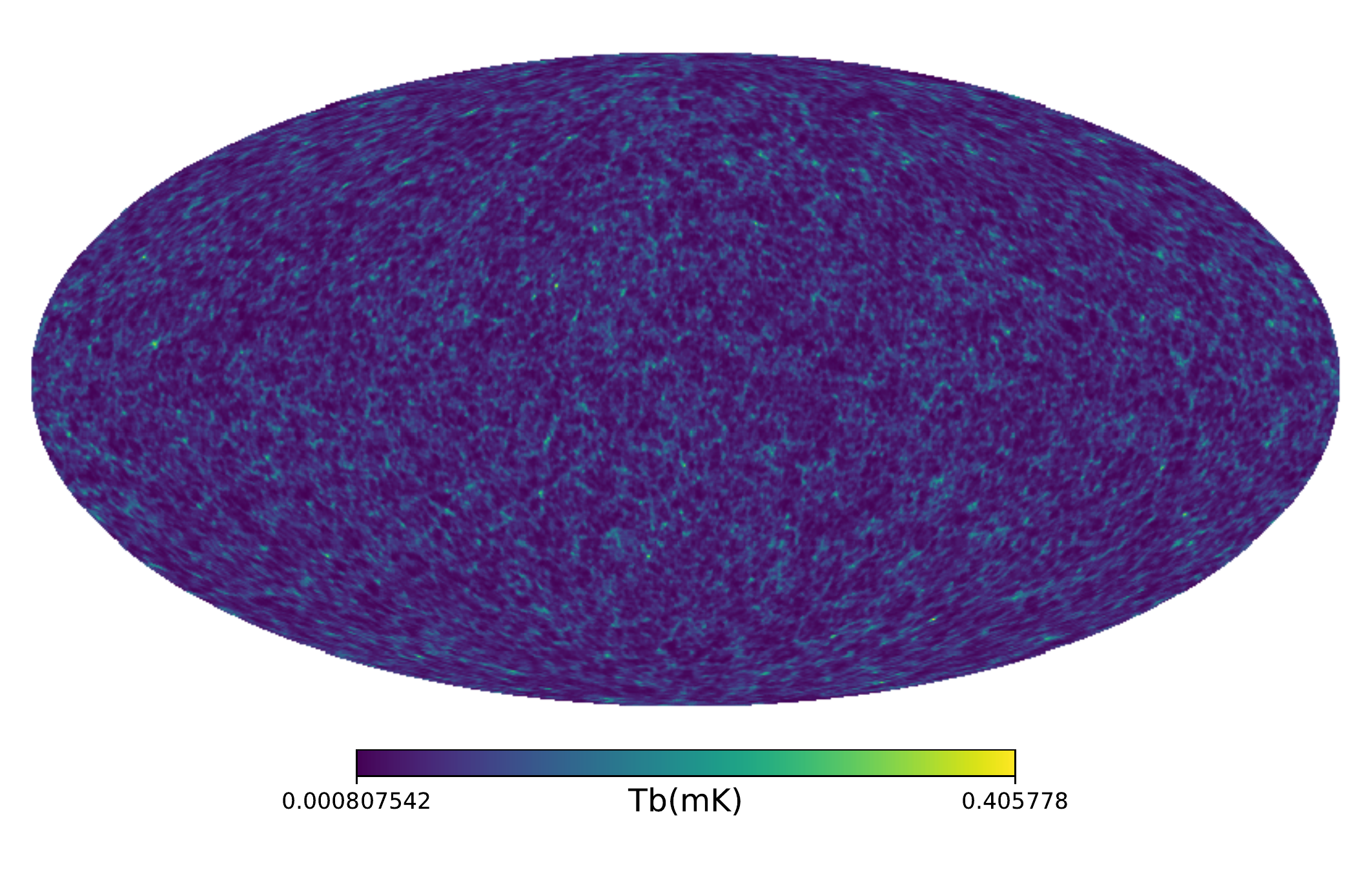}
    \caption{Mollview of brightness temperature map using abundance matching method to populate \hi\ gas in galaxies in real space from 990 MHz to 1000 MHz.}
    \label{fig:ammap}
\end{figure}

Halo Abundance Matching (HAM) and Sub-Halo Abundance Matching (SHAM) have been widely used to link the dark matter halos to galaxies in observations. Under the assumption that more luminous galaxies (with more \hi\ mass) are populated inside more massive halos (sub-halo), we can populate the halos in the simulations with galaxies, which is adopted in HR4 galaxy catalog generation \citep{hong2016halogalaxy}. In HR4 galaxy catalog, the position, velocity and an "artificial mass-like quantity", which is also used to estimate the galaxy luminosity with abundance matching, are provided. We use the Schechter form \hi\ mass function measured from the ALFALFA catalog \citep{Martin2010ALFALFA},
\begin{equation}
\begin{aligned}
    &\phi(m_{\rm \hi\,})d\log_{10}\left(m_{\rm \hi\,}\right)=\ln(10)\phi_*\left(\dfrac{m_{\rm \hi\,}}{m^{*}}\right)^{\alpha+1}\exp{\left(-\dfrac{m_{\rm \hi\,}}{m_{*}}\right)}d\log_{10}(m_{\rm \hi\,})\,,\\
    &\phi_* = 0.0048\\
    &\alpha = -1.33\\
    &m_*=10^{9.96}\left(\dfrac{0.7}{h}\right)^2\,,
\end{aligned}
\end{equation}
where $m_{\rm \hi\,}$ is in units of $M_{\odot}$. We also calculated the abundance of galaxy "mass" in the redshift range $0.127<z<0.449$ in the HR4 galaxy light cone catalog. By matching the abundance of galaxy "mass" and the \hi\ mass function, we get a \hi\ mass-galaxy "mass" relation. Therefore, we can populate \hi\ mass into the galaxy catalog. Notice that the measurement of \hi\ mass function from ALFALFA is at $z=0$, which is not compatible with the light cone catalog. It is just an approximated way of populating \hi\ mass into galaxies. Then we sliced the catalog into redshift bins of 10 MHz each. We calculate the density of \hi\ in each redshift using {\tt HEALPix} pixelization \citep{healpix} with $N_{\rm side}=256$. Finally, using Eq. (\ref{eq:Tb}), we calculate the brightness temperature distribution. This set of mock map is labelled as HAM mock map in the following sections.

In Fig.  \ref{fig:ammap}, we show the mollview of brightness temperature map at 990 MHz - 1000 MHz bin, in units of mK, as an example. The map represents the \hi\ distribution in real space.
\subsection{HOD Populate}\label{subsec:hipop}
\begin{figure}
    \centering
    \includegraphics[width=0.45\textwidth]{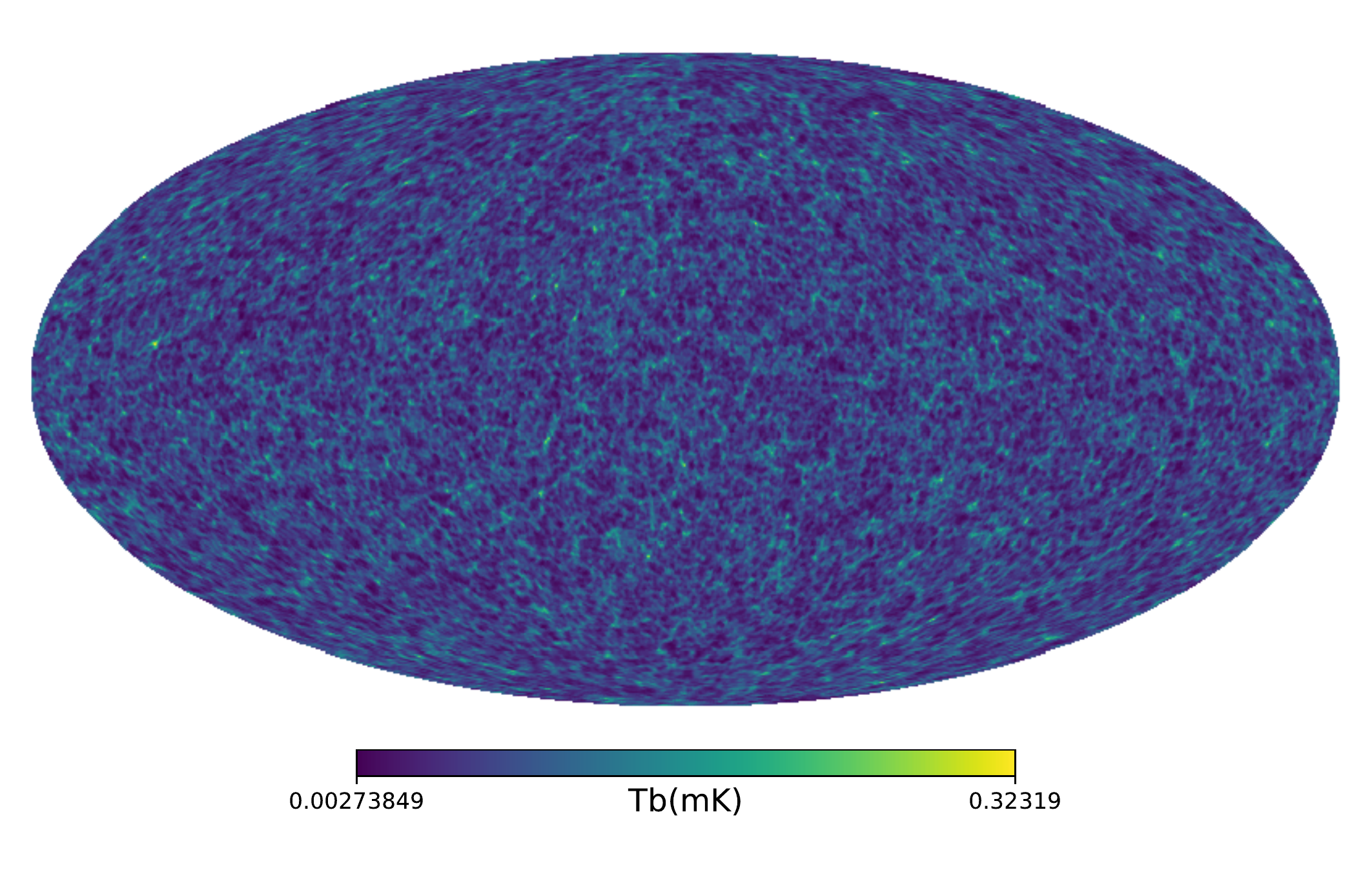}
    \caption{Mollview of brightness temperature map using HOD method to populate \hi\ gas in halos in real space from 990 MHz to 1000 MHz.}
    \label{fig:hodmap}
\end{figure}

Using abundance matching method has the following shortage,
\begin{itemize}
    \item[1] Use the \hi\ mass function at $z=0$ to match all galaxies at different redshift is not fair, the evolution with redshift is ignored and the \hi\ mass is mismatched,
    \item[2] The assumption that galaxy with more "mass" has more \hi\ mass may not hold,
    \item[3] It is not convenient to calculate the bias from theory to compare with the mock map,
    \item[4] If there is only halo catalog, this abundance matching method do not apply, while many fast simulations only provide halo catalog.
\end{itemize}
Therefore, we need to use the HOD approach to obtain the mock map. 

Using the FoF halo light cone catalog from HR4 simulation, we can use the \hi\ mass-halo mass relation described by Eq. (\ref{eq:hihalofit}) to populate \hi\ mass into the halo catalog. Then we follow the same process described in the last subsection, separate the catalog into 30 bins and pixelize the full sky map using {\tt HEALPix} with $N_{\rm side}=256$. We get the brightness temperature map and show it in Fig. \ref{fig:hodmap}. This mollview map is from 990 MHz to 1000 MHz, representing the real space distribution of \hi\ gas, the same as Fig.  \ref{fig:ammap}. Comparing these two maps, we can see that different ways of populating \hi\ gas into galaxies (halos) leads to different results. Though the map looks very similar at large scale, the brightness temperature fluctuation in the HAM map is much higher than that in the HOD map. This illustrates different bias of \hi\ distribution. With abundance matching method, \hi\ gas is more clustered than that in the HOD method. More detailed discussion will be provided in Sec. \ref{sec:result}. With intensity mapping, we can tell different ways of \hi\ clustering and occupation in halos.

\subsection{Redshift Space Distortion}\label{subsec:rsd}
\begin{figure}
    \centering
    \includegraphics[width=0.45\textwidth]{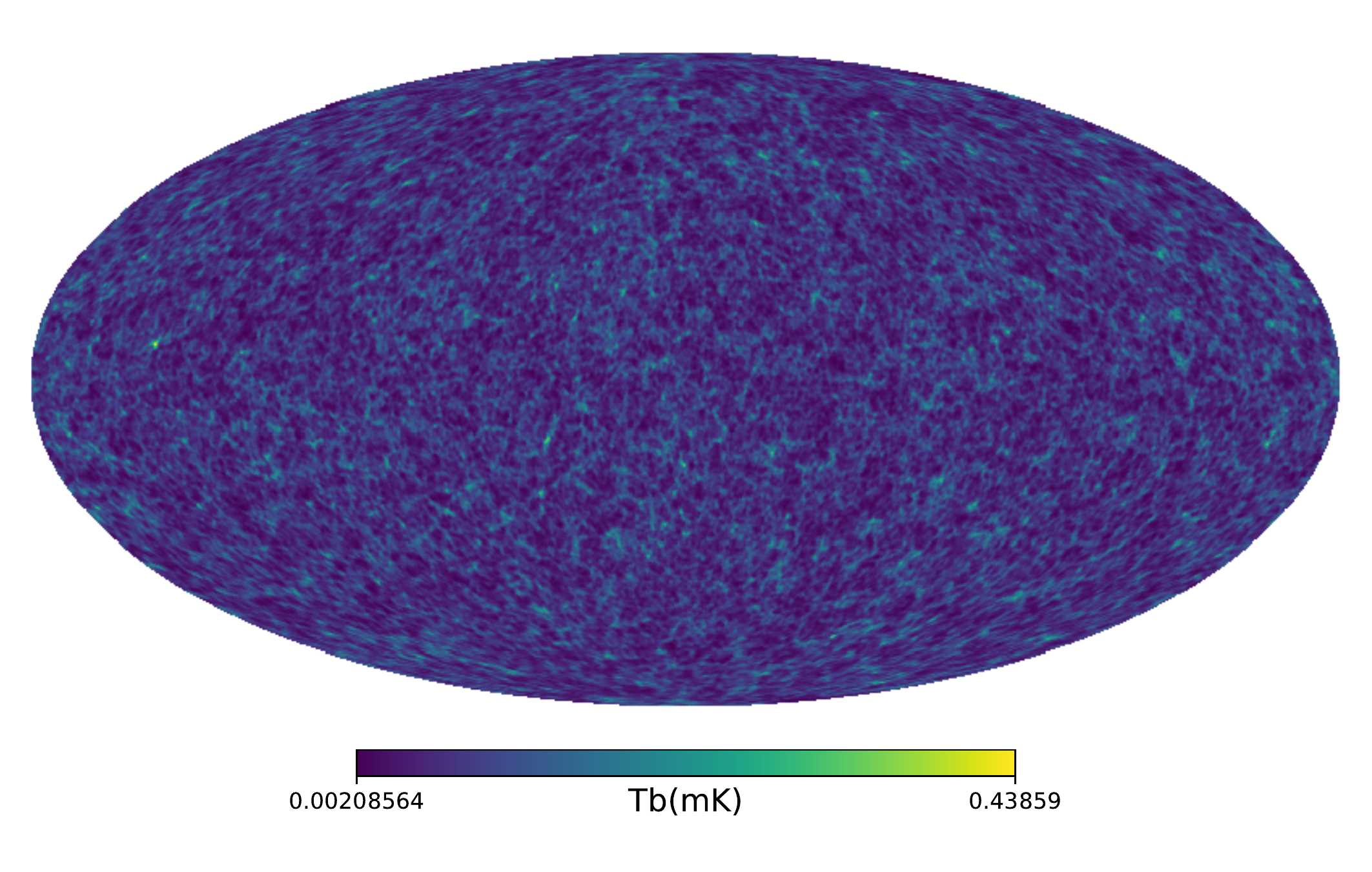}
    \includegraphics[width=0.45\textwidth]{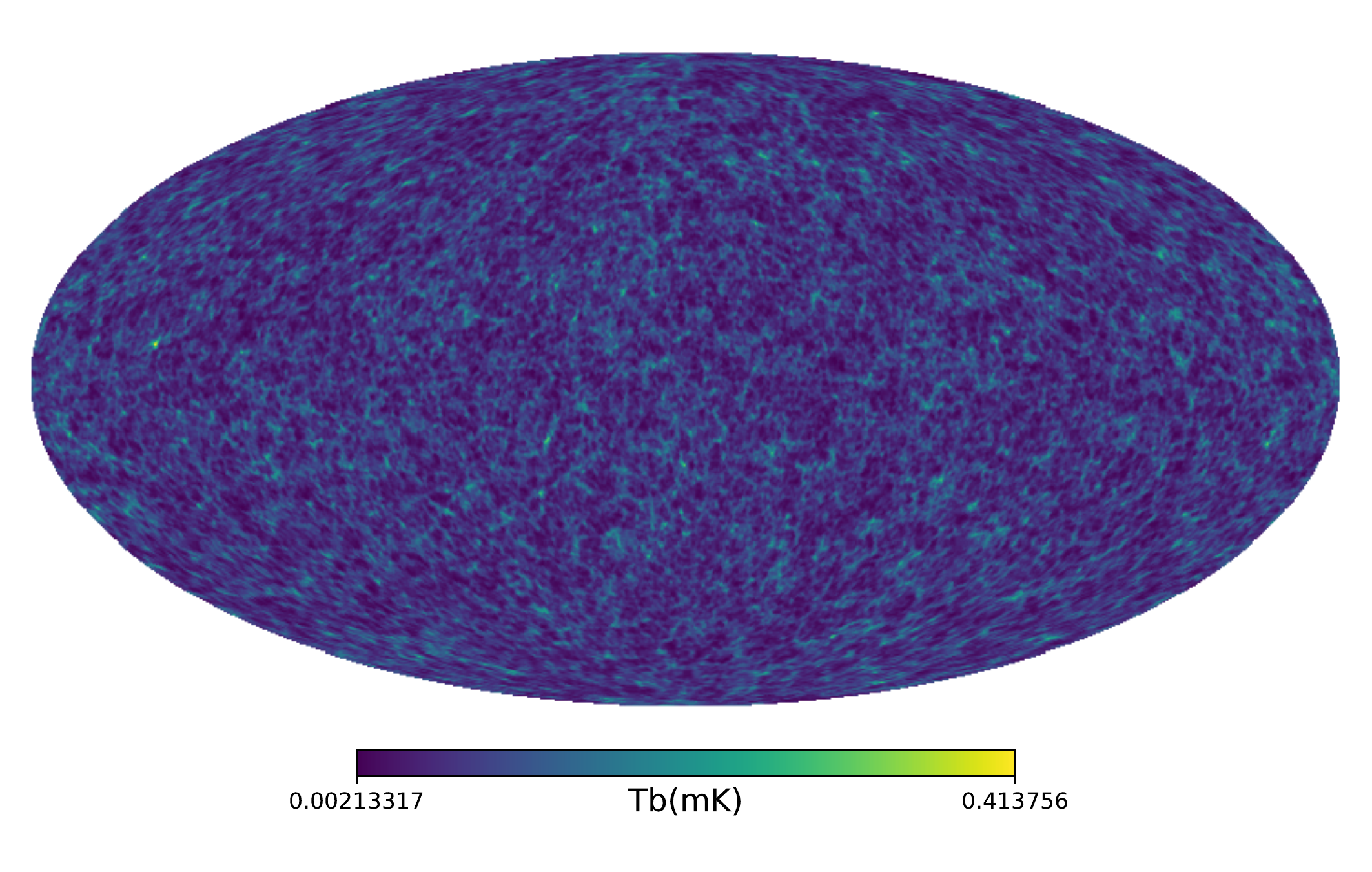}
    \caption{Mollview of brightness temperature map using HOD method to populate \hi\ gas in halos in redshift space from 990 MHz to 1000 MHz. We have only considered the velocity of halos as point source in the upper panel, which mainly only contains Kaiser effect. In the lower panel, we have considered the velocity dispersion of \hi\ gas in sources, which means the FoG effect is taken into account.}
    \label{fig:rsdmap}
\end{figure}

The intensity map we observe is binned according to the frequency, which is determined by the redshift of the 21-cm emission source. Thus, we are observing in redshift space rather than real space. We need to consider not only the redshift of emission sources due to the Hubble flow, but also consider the redshift contribution from the peculiar velocity of the sources. The positions of the 21-cm emission sources are different in redshift space from that in real space. This is known as RSD effect, which consists of Kaiser effect in large scale \citep{kaiser1987} and Finger-of-God (FoG) effect in small scale \citep{FoG}. We considered two possible sources of the RSD effect in the mock map,
\begin{itemize}
    \item[1] The peculiar velocity of galaxies (halos) as a point source, which mainly contributes to Kaiser effect,
    \item[2] The velocity dispersion inside the halos, which mainly contributes to FoG effect.
\end{itemize}

In Fig.  \ref{fig:rsdmap}, we show the mollview mock map of the brightness temperature adopting the point source assumption in the upper panel, while the result of additionally considering velocity dispersion is shown in the lower panel. Under the point source assumption it is easy to adopt the RSD effect, since the redshift of the source can be calculated by,
\begin{equation}
    z' = (1+z)(1+v_{r}/c)-1\,,
\end{equation}
where $z$ is the redshift of the point source calculated according to its distance to the observer and hubble flow, which is provided in the light cone catalog, $v_{r}$ is the peculiar velocity along the line-of-sight direction and $c$ is the speed of light. Then we can bin the sources according to their redshift $z'$, calculate the \hi\ density with $N_{\rm side}=256$ {\tt HEALPix} pixelization and calculate the corresponding brightness temperature. Comparing to the real space mock map, after considering the RSD effect, some of the sources are moved away from its real space frequency range, and some of them are moved into the frequency bin. Thus, we can see slightly different large scale structures between the real space mock map and the redshift space mock map. 

The mock map considering velocity dispersion of \hi\ gas in halos is built following the steps,
\begin{itemize}
    \item[1] Calculate the velocity dispersion of \hi\ gas in each halo according to the halo mass $m$ and real redshift $z$,
    \begin{equation}
        \sigma_v(z)=(4.9z+30)\left(\dfrac{m}{10^{10}M_{\odot}/h}\right)^{(0.0091z+0.36)},
    \end{equation}
    which is fitted from IllustrisTNG simulation \citep{F18} and adopted in \citet{zhang2020parameter} (notice that this is a fitting formula using multiple data points at different redshifts, the calculated result at $z=0$ is not exactly the same as Eq. \ref{eq:sigmav0}),
    \item[2] Calculate the corresponding comoving distance variance $\delta_s = \dfrac{1+z}{H(z)}\dfrac{\sigma_v(z)}{\sqrt{3}}$\,,
    \item[3] For a given redshift bin from $z_1$ to $z_2$, where $z_1<z_2$, calculate the corresponding comoving distance range from $s_1$ to $s_2$, select the halos within the range $s_1-3\delta_s$ and $s_2+3\delta_s$ in redshift space,
    \item[4] Calculate the \hi\ mass contribution $w$ from the selected halos in this given bin, for a given halo at $s_h$,
    \begin{equation}
        \begin{aligned}
        &t_1 = (s_h-s_1)/\delta_s\,,\\
        &t_2 = (s_h-s_2)/\delta_s\,,\\
        &w = \dfrac{m_{\rm \hi\,}}{\sqrt{2\pi}}\int_{t_1}^{t_2}e^{-t^2/2}dt\,,
        \end{aligned}
    \end{equation}
    \item[5] Calculate the \hi\ density in this bin and the related brightness temperature.
\end{itemize}
By doing so, we have assumed that the mass distribution of \hi\ in one halo in redshift space follows Gaussian distribution. This process can provide a rough estimation of FoG effect in the mock map.

\subsection{Rescale the Map}\label{subsec:rescale}

The average brightness temperature $\bar{T}_b$ can be calculated from the mock maps at different redshifts, it also represents the \hi\ abundance $\Omega_{\rm \hi\,}$ since the mean temperature is proportional to it. The relation between $\bar{T}_b$ and $\Omega_{\rm \hi\,}$ is
\begin{equation}
    \bar{T}_b=189h\dfrac{H_0(1+z)^2}{H(z)}\Omega_{\rm \hi\,}\text{mK}\,.
\end{equation}
 In the mock map, since we have neglected the contribution from halos whose mass is smaller than the resolution, we may have missed some of the \hi\ gas in small halos. On the other hand, we may have also overestimated the \hi\ mass in halos comparing to the real observation by various reasons, such as overestimating the \hi\ mass-halo mass relation. Therefore we need to rescale the maps to meet the requirement of $\Omega_{\rm \hi\,}$. We can rescale the maps according to the required $\Omega_{\rm \hi\,}$ value, by applying $T'_b=T_b\dfrac{\Omega'_{\rm \hi\,}}{\Omega_{\rm \hi\,}}$.

\section{Results of the Mock}\label{sec:result}
\subsection{Angular Power Spectra}\label{subsec:cls}

We would like to address three questions in this subsection,
\begin{itemize}
    \item[1] What is the effect of different ways of linking \hi\ gas with galaxies (halos)? 
    \item[2] What is the effect of RSD in angular power spectrum?
    \item[3] Can we identify the above difference with BINGO telescope?
\end{itemize}

\begin{figure*}
    \centering
    \includegraphics[width=0.33\textwidth]{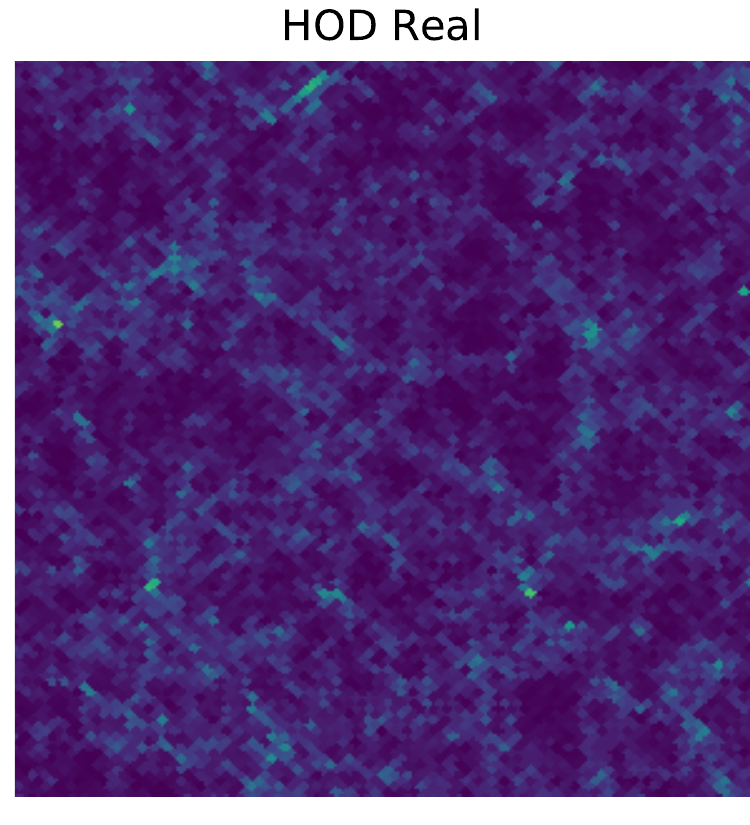}
    \includegraphics[width=0.33\textwidth]{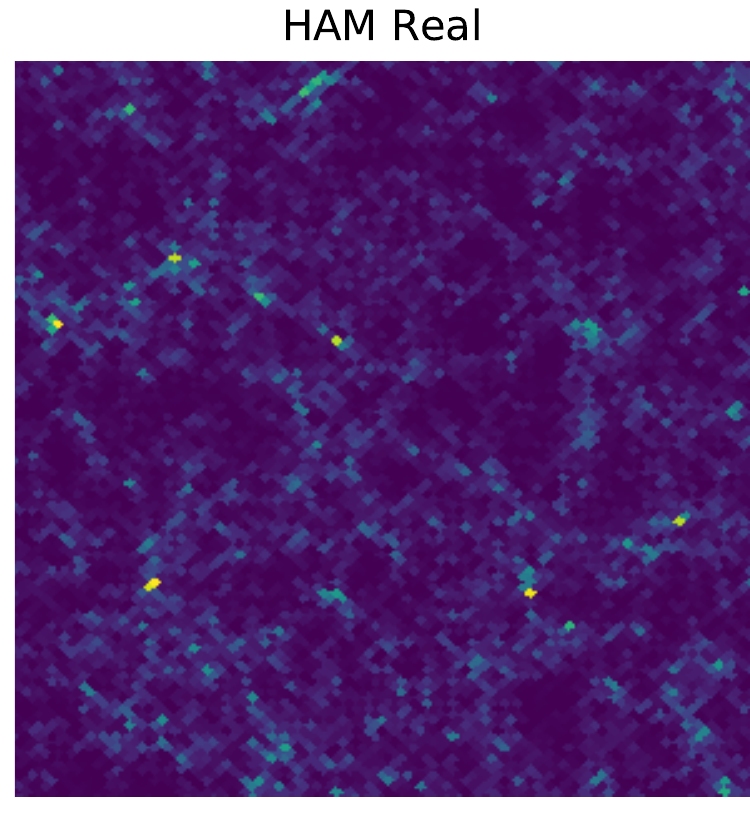}
    \includegraphics[width=0.33\textwidth]{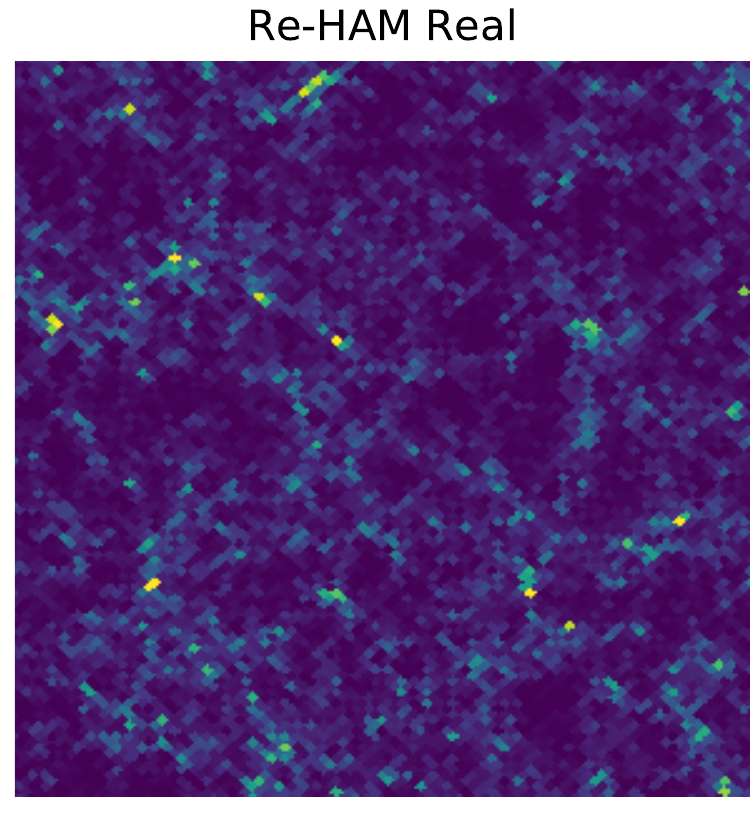}
    \includegraphics[width=0.33\textwidth]{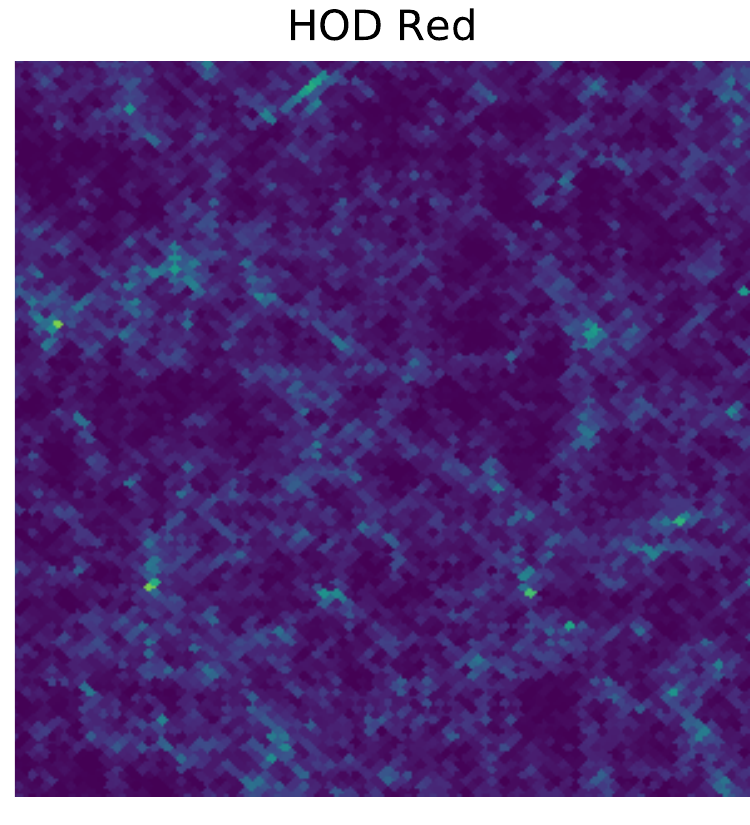}
    \includegraphics[width=0.33\textwidth]{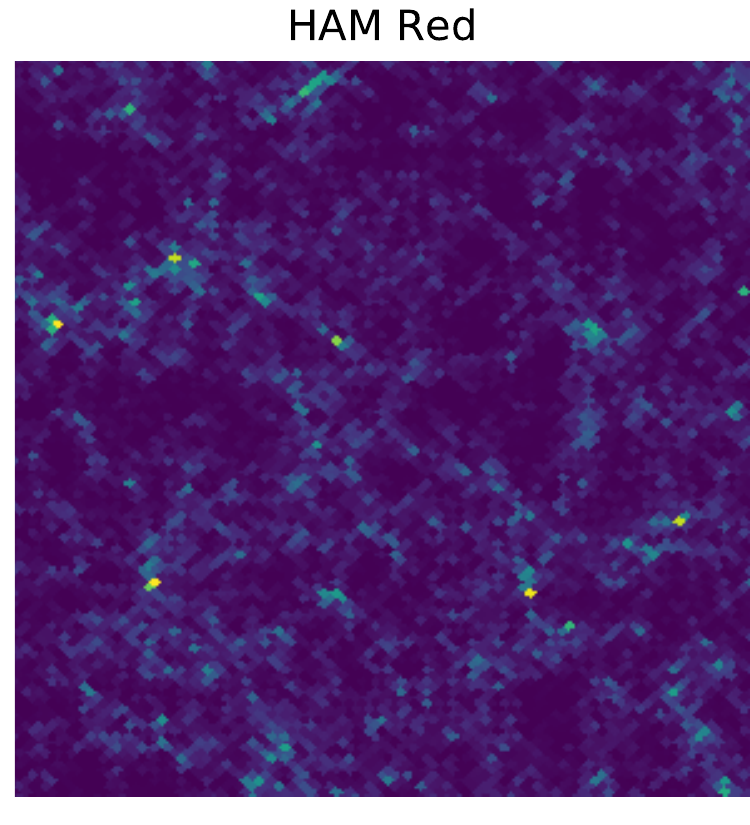}
    \includegraphics[width=0.33\textwidth]{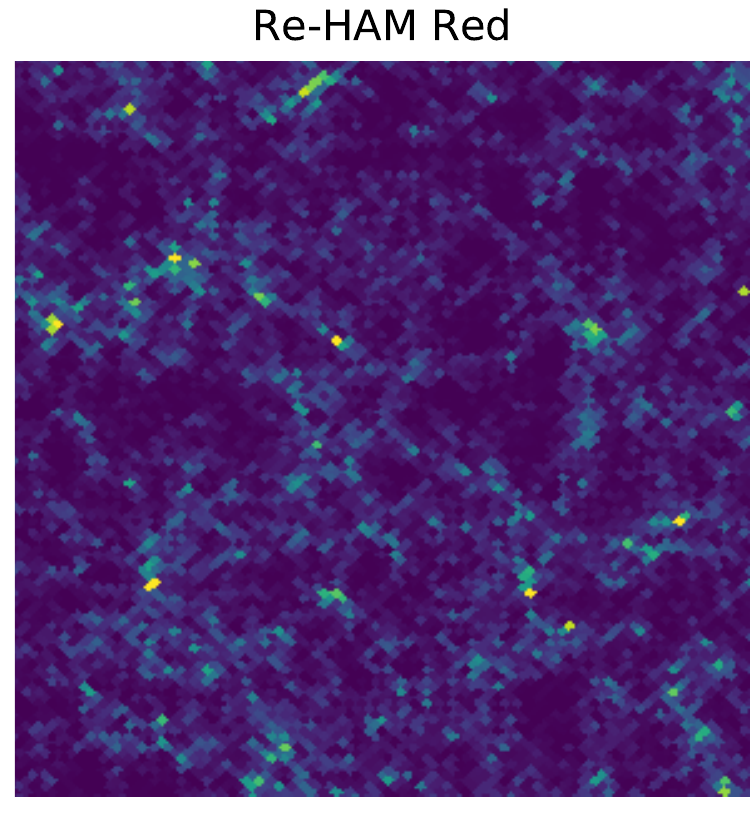}
    \includegraphics[width=0.33\textwidth]{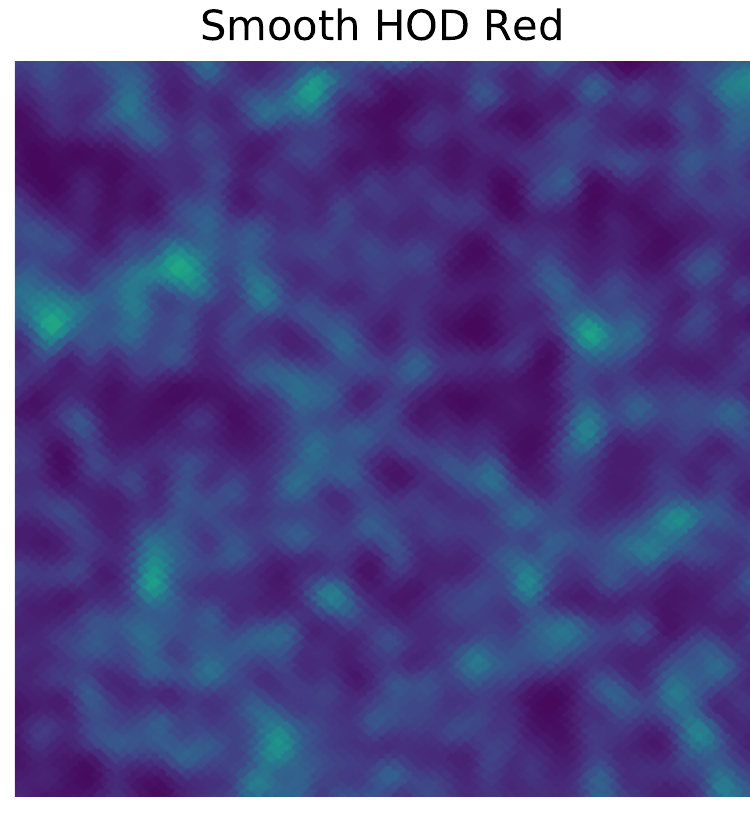}
    \includegraphics[width=0.33\textwidth]{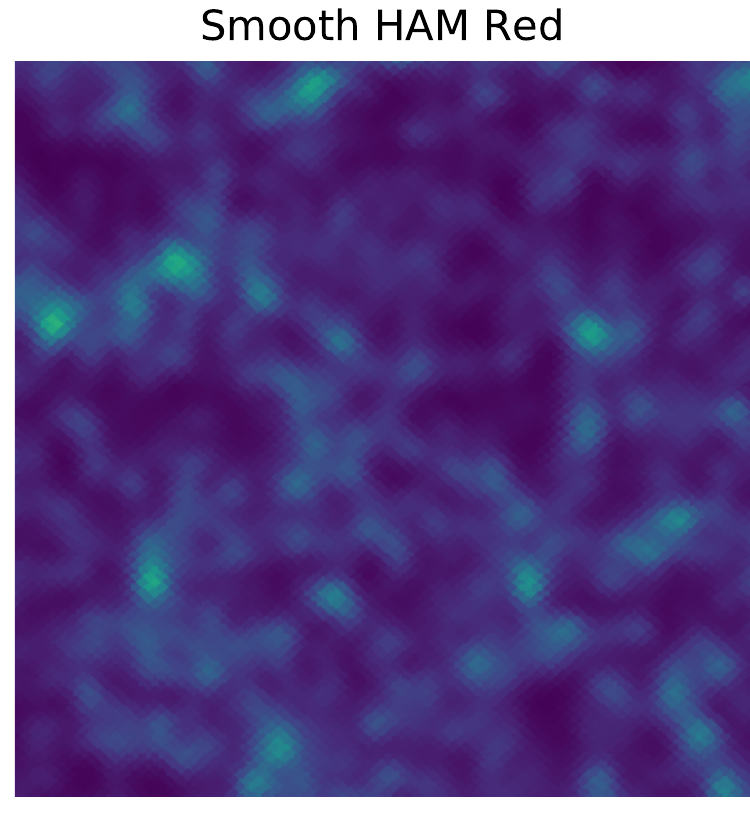}
    \includegraphics[width=0.33\textwidth]{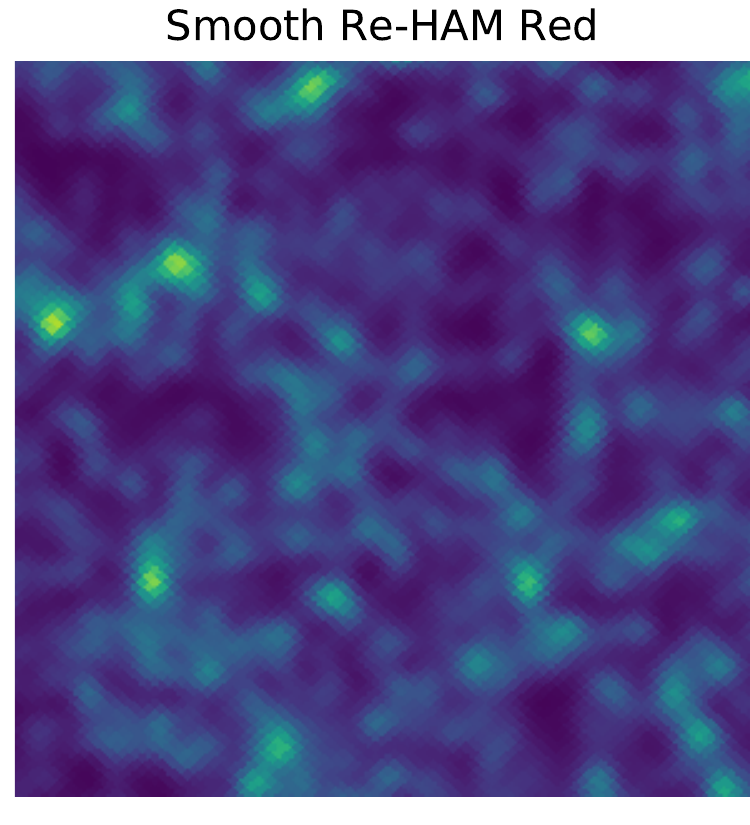}
    \caption{Zoom-in view of the central $10\times10$ degree$^2$ brightness temperature distribution of the mock map from 990 MHz to 1000 MHz.}
    \label{fig:cartmap}
\end{figure*}

In Fig.  \ref{fig:cartmap}, we have shown the zoom-in view of a $10\times10$ degree$^2$ brightness temperature map of the full sky mock map from 990 MHz to 1000 MHz. We have compared the map generated using HOD method (labeled as HOD), abundance matching method (labeled as HAM) and re-scale the HAM map to the same $\Omega_{\rm \hi\,}$ as the HOD map (labeled as Re-HAM). We have also shown the map smoothed by a Gaussian kernel with FWHM = 40 arcmin.The same color map has been given in each row for a fair comparison. The $T_b$ contrast in HAM map is clearly much higher than that in HOD map, which can be more easily identified in the Re-HAM map. In the HAM map, there are more \hi\ gas inside clusters than in the HOD map. Since the bias of more massive halos is higher than less massive halos, we will expect that with the same $\Omega_{\rm \hi\,}$, the angular power spectrum of the mock map generated by abundance matching method will be higher than the map generated using HOD. 
\begin{figure}
    \centering
    \includegraphics[width=0.45\textwidth]{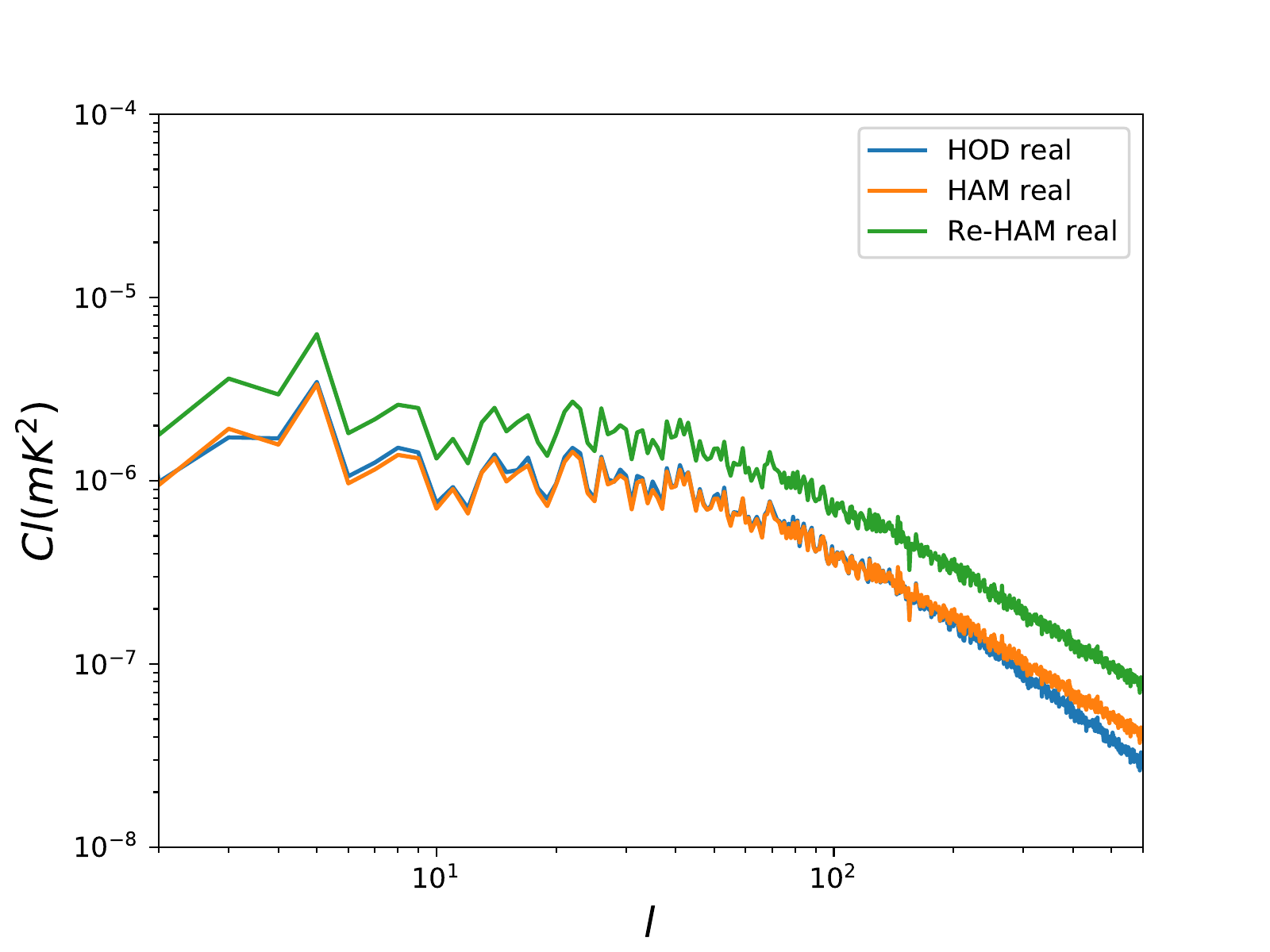}
    \caption{The comparison of angular power spectrum among HOD mock, HAM mock and Re-HAM mock in real space. With the same $\Omega_{\rm \hi\,}$, Re-HAM mock has higher angular power spectrum than the HOD mock, which means it has a higher bias. Though very close at large scale, HAM mock has higher angular power spectrum than the HOD mock at small scale due to the existence of substructures in HAM mock.}
    \label{fig:hod_vs_ham}
\end{figure}

In Fig. \ref{fig:hod_vs_ham}, our expectation is verified. Re-HAM and HOD share the same $\Omega_{\rm \hi\,}$, however the angular power spectrum of the Re-HAM mock map is much higher than the HOD mock map. The HAM mock shows very close angular power spectrum to the HOD mock by coincidence. We can also notice that at large $l$ ($l>200$), the real space angular power spectrum of the HAM mock is slighter larger than the HOD mock. This phenomena can also be noticed in Fig. \ref{fig:hodpk}. We know that in the HOD mock, the \hi\ gas are assumed to locate at the center of the halo, without any substructures. 
While in the mocks that contain subhalos, the subhalo distribution inside main halos leads to additional clustering at small scale, which is known as one-halo term. This can be seen when comparing the HR4 HAM mock and HR4 HOD mock in Fig. \ref{fig:hod_vs_ham}. It can also be seen when comparing ELUCID SAM mock and ELUCID HOD mock in Fig. \ref{fig:hodpk}. Therefore, it is expected that at small scale, the real space power spectrum of HOD mock has smaller amplitude than that of the HAM mock or SAM mock in real space. However, when considering RSD effect, since the subhalos in halos are also moving randomly in the frame of the center of mass of the main halo, there will be FoG effect that can suppress the power spectrum at small scale. Thus, in the "red" mock, the higher power from one-halo term and the lower power from FoG counter each other, the power spectrum between HOD mock and SAM from ELUCID simulation are getting closer. Furthermore, in the "red2" mock, the FoG effect is taken into account in the HOD mock, the difference between HOD and SAM are revealed again. 
\begin{figure}
    \centering
    \includegraphics[width=0.45\textwidth]{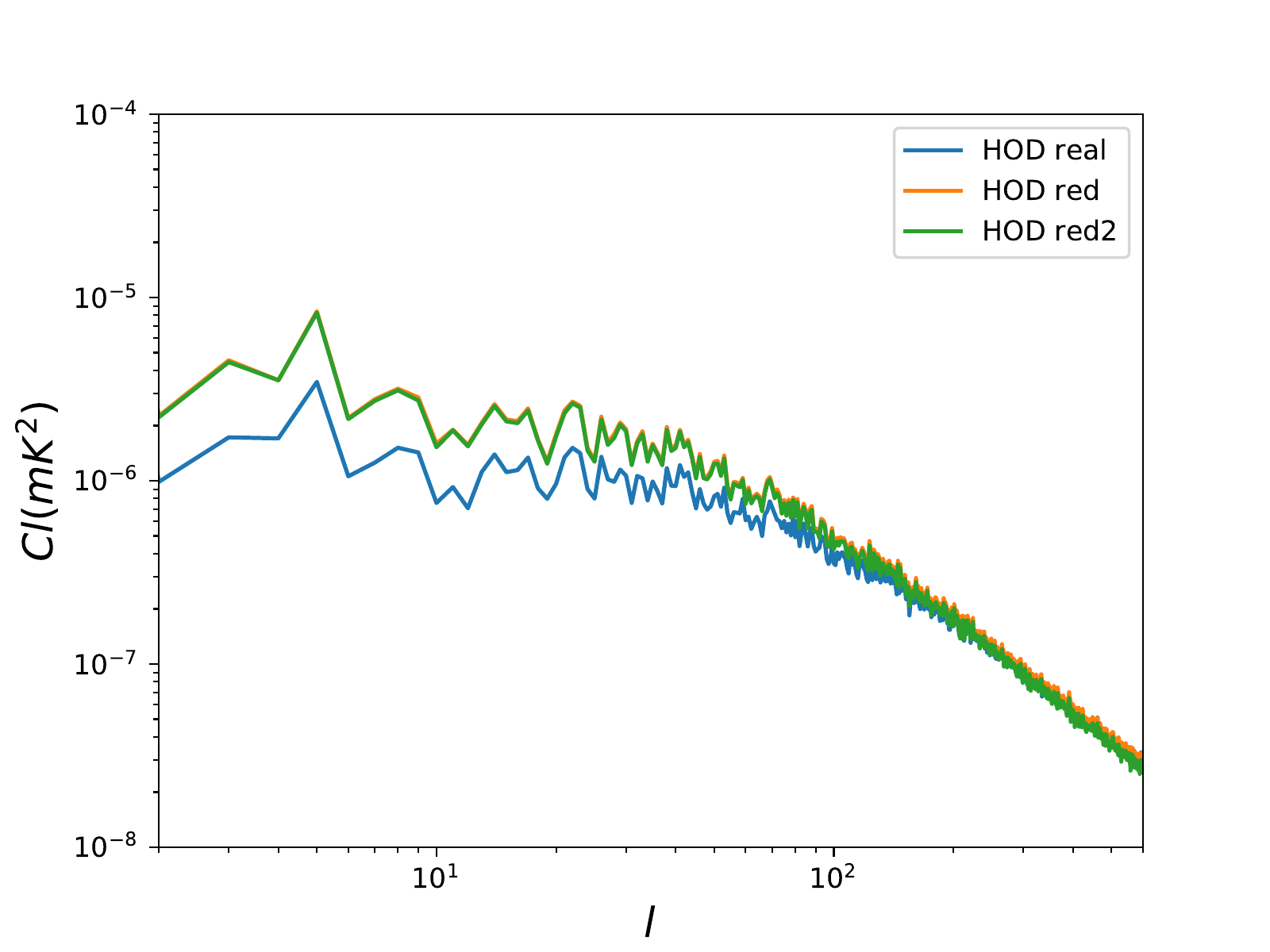}
    \caption{The comparison of angular power spectrum between the HOD mock in real space and redshift space. The power in redshift space is higher than that in real space at large scale due to Kaiser effect. No matter we treat the \hi\ gas in halos as point sources ("red") or extended sources due to internal velocity dispersion ("red2") in redshift space, the difference between them is not significant.}
    \label{fig:red_vs_red2}
\end{figure}

In Fig.  \ref{fig:red_vs_red2}, we have shown the comparison between the HOD real space mock (labeled as "real"), the HOD redshift space point source mock (labeled as "red") and HOD redshift FoG effect included mock (labeled as "red2"). The angular power spectrum difference between the "real" mock and "red" mock is clear at large scale due to Kaiser effect, but the difference between "red" mock and "red2" mock due to FoG effect at small scale is not significant. This is because of the bin width of the mock being 10 MHz, that corresponds to more than $20 \text{ Mpc}/h$, while the typical velocity dispersion is no larger than $1000 \text{ km/s}$, which roughly corresponds to $10 \text{ Mpc}/h$ in redshift space. Therefore, the power spectrum suppression introduced by the velocity dispersion is averaged out by the large bin width. 
\begin{figure}
    \centering
    \includegraphics[width=0.45\textwidth]{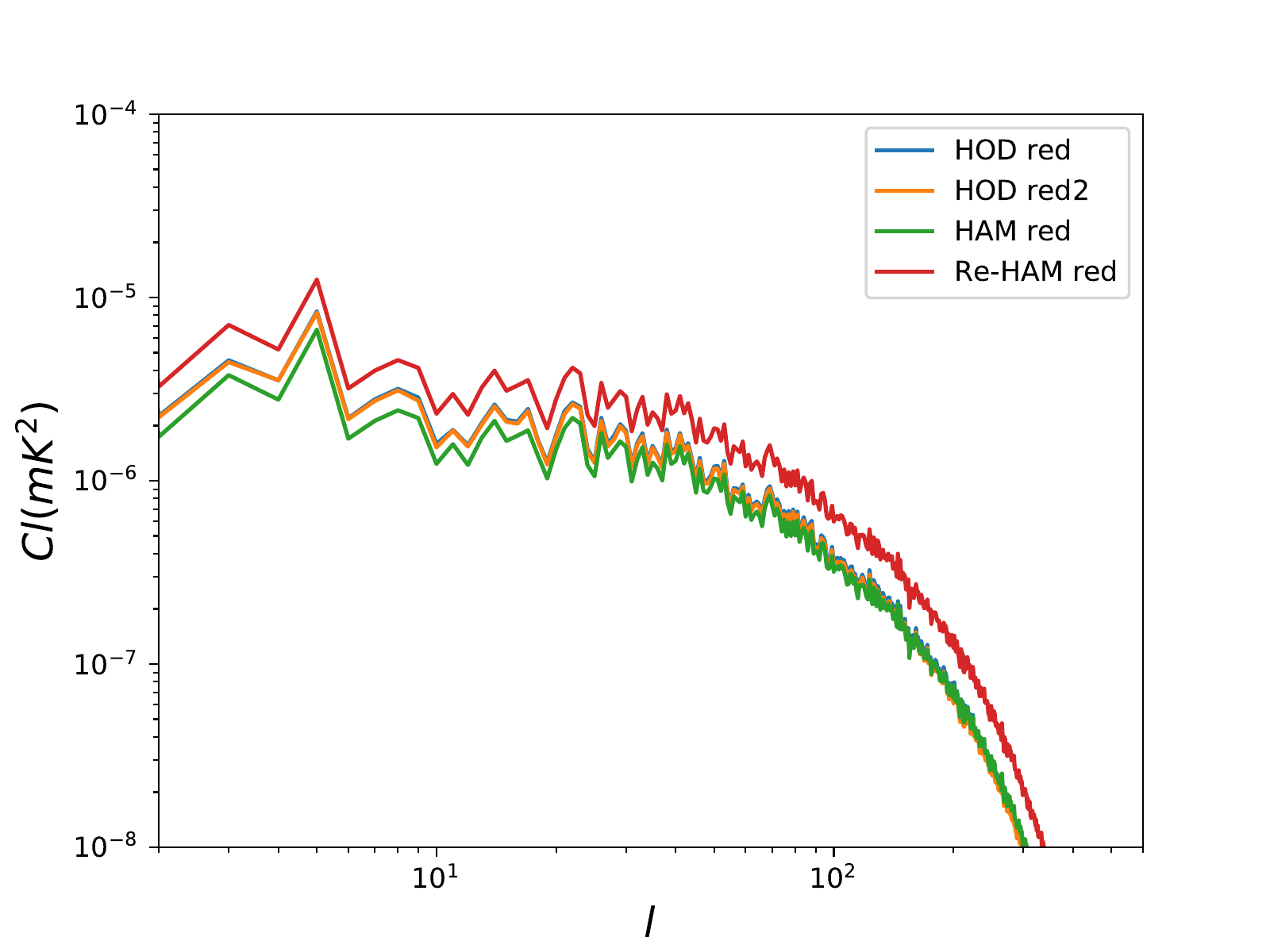}
    \caption{We show the angular power spectrum of the HOD mock, HAM mock and Re-HAM mock in redshift space. Re-HAM mock is clearly higher due to a higher bias. The difference between HAM mock and HOD mock due to one-halo term and the difference between "red" mock and "red2" mock due to FoG effect at small scale are not significant because of smoothing introduced by Gaussian kernel FWHM = 40 arcmin.}
    \label{fig:smoothcl}
\end{figure}

In Fig. \ref{fig:smoothcl}, we have compared the angular power spectrum of the mocks smoothed by Gaussian kernel FWHM = 40 arcmin, which is the beam size of BINGO. After smoothing, the difference at small scale is easily covered. The difference of bias can still be clearly identified by comparing HOD mock and Re-HAM mock, which share the same $\Omega_{\rm \hi\,}$ value. Therefore, we can answer the questions raised in the beginning of this subsection,
\begin{itemize}
    \item[1] Different ways of linking \hi\ gas with galaxies (halos) lead to different bias and one-halo term, the bias difference can be seen in the total power of angular power spectrum and the difference of one-halo term can be seen in the small scale,
    \item[2] Kaiser effect leads to higher angular power spectrum at large scale. FoG effect leads to suppression of power spectrum at small scale, but cannot be discriminated due to 10 MHz bin size,
    \item[3] Gaussian kernel smoothing eliminates the difference at small scale, but the bias difference shown in all scales and the Kaiser effect shown in large scales can still be easily identified.
\end{itemize}

\subsection{Pixel Histogram}\label{subsec:hist}
\begin{figure}
    \centering
    \includegraphics[width=0.45\textwidth]{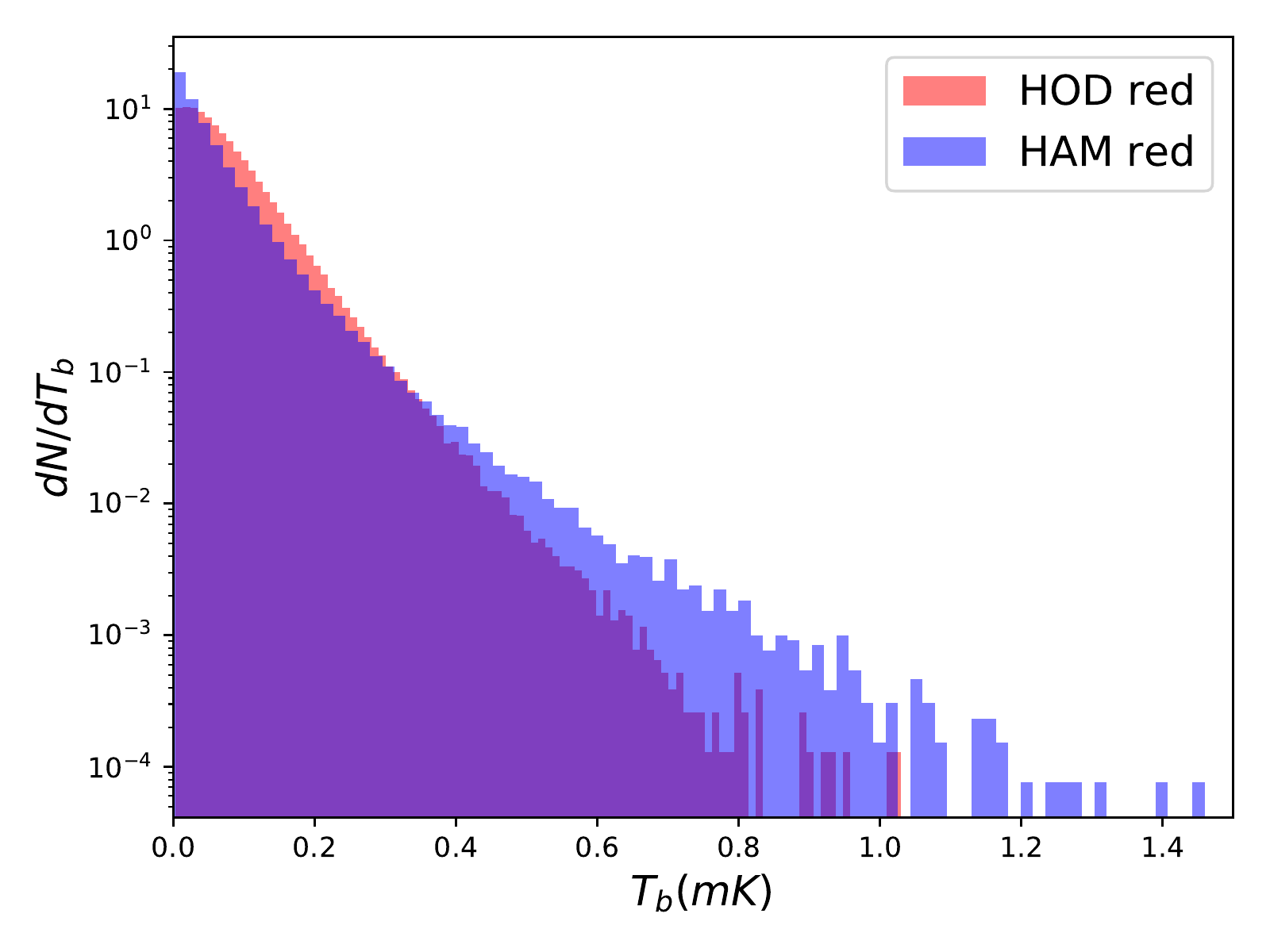}
    \includegraphics[width=0.45\textwidth]{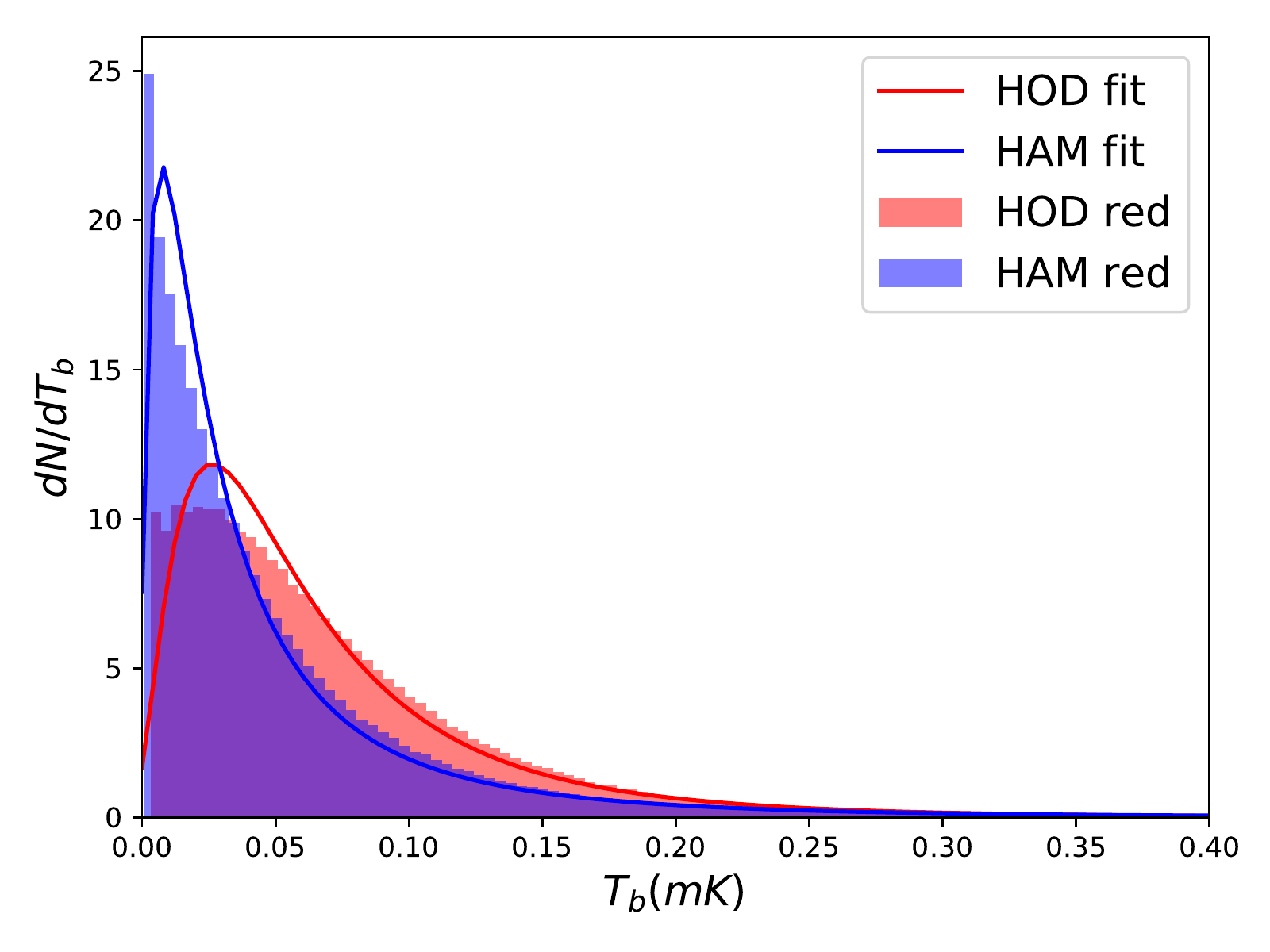}
    \caption{The pixel histogram of the brightness temperature distribution of the mock map from 990 MHz to 1000 MHz in redshift space. In the upper panel, we show the histogram comparison of the HOD map and the HAM map from 0 mK to 1.5 mK. In the lower panel, we zoom in to the range from 0 mK to 0.4 mK. The lognormal fitted curves are also shown. The HAM map has clearly more extremely high brightness temperature pixels, as expected by looking at Fig. \ref{fig:cartmap}.}
    \label{fig:hist}
\end{figure}
We have shown that the brightness temperature fluctuation in the HAM mock is larger than in the HOD mock in Fig.  \ref{fig:cartmap}. More quantitatively, we calculated the histogram of the pixels in the unsmoothed mock map. In the upper panel of Fig.  \ref{fig:hist}, we showed the histogram of the pixels, notice that we have excluded all the pixels which are zero. As expected, there are more extremely high brightness temperature pixels in the HAM mock than in the HOD mock. 

On the other hand, the lognormal distribution was widely used to simulate dark matter tracer fields \citep{Xavier:2016}, which is the 21-cm intensity map in our case. We calculated the histogram of the pixels from 0 mK to 0.4 mK and fitted with the lognormal distribution using the stats routine in scipy \citep{scipy}. The result is shown in the lower panel of Fig. \ref{fig:hist}. We found that the lognormal distribution can reasonably represent the measured results from our mock map. This means that using lognormal distribution to simulate the brightness temperature map is reasonable, but with clear limitations as well.

The difference between the pixel histogram of the HOD mock and the HAM mock shows that the way how the \hi\ gas is distributed inside halos can largely affect the intensity map, not only in the angular power spectrum, but also in other statistical measurements. 

\subsection{Lognormal Fitting}\label{subsec:lognorm}
\begin{figure*}
\centering
\includegraphics[width=0.45\textwidth]{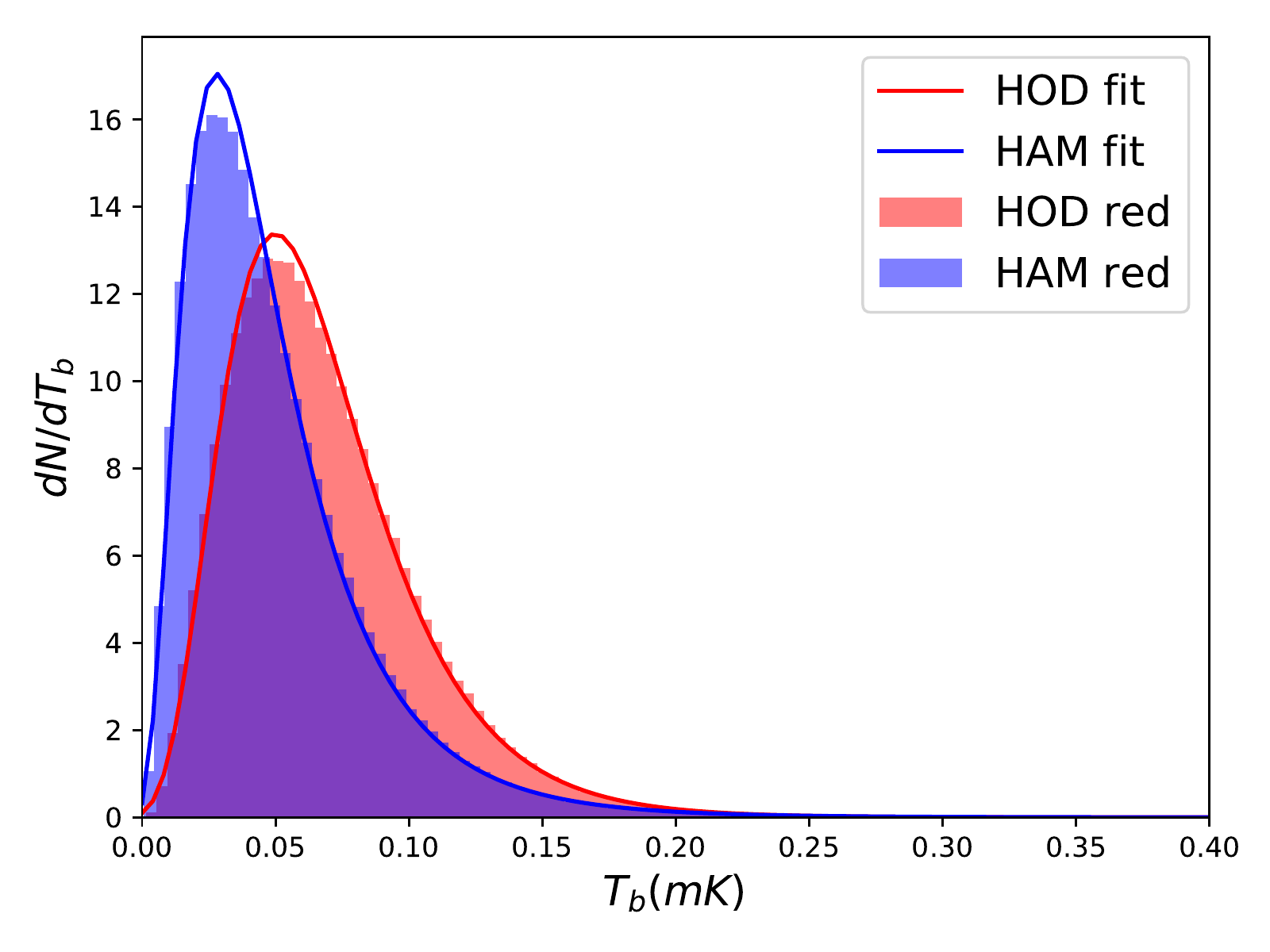}
\includegraphics[width=0.45\textwidth]{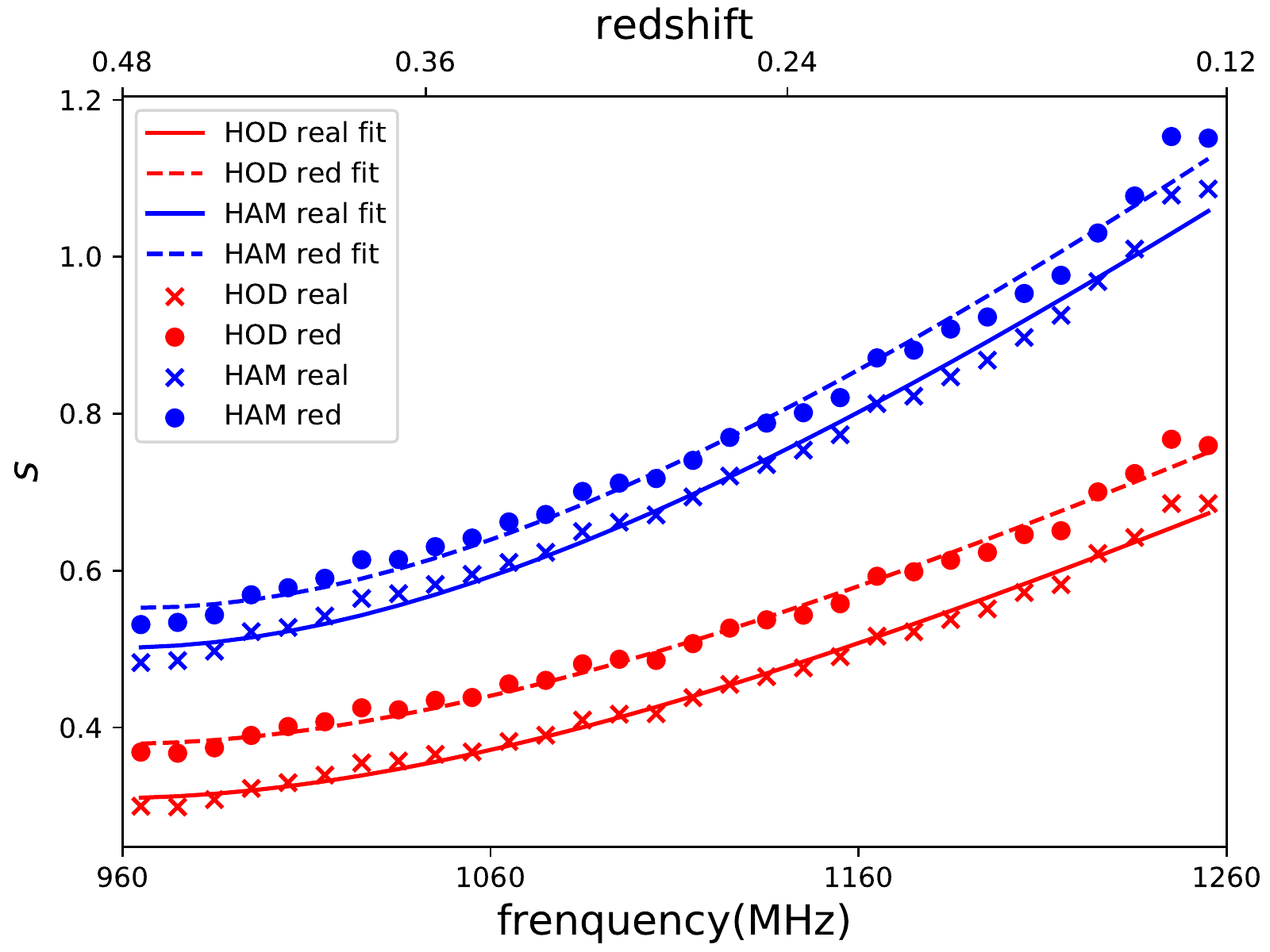}
\includegraphics[width=0.45\textwidth]{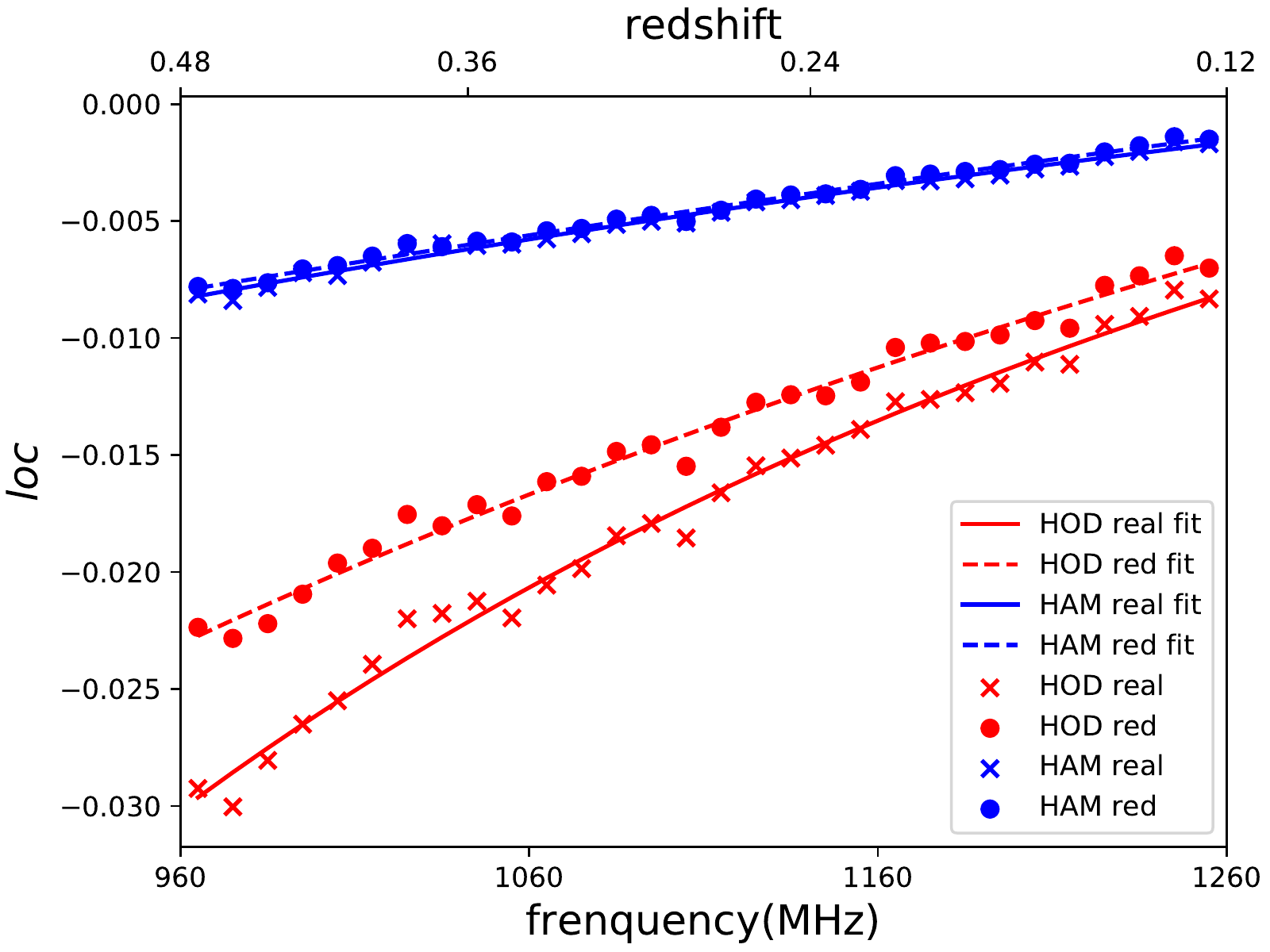}
\includegraphics[width=0.45\textwidth]{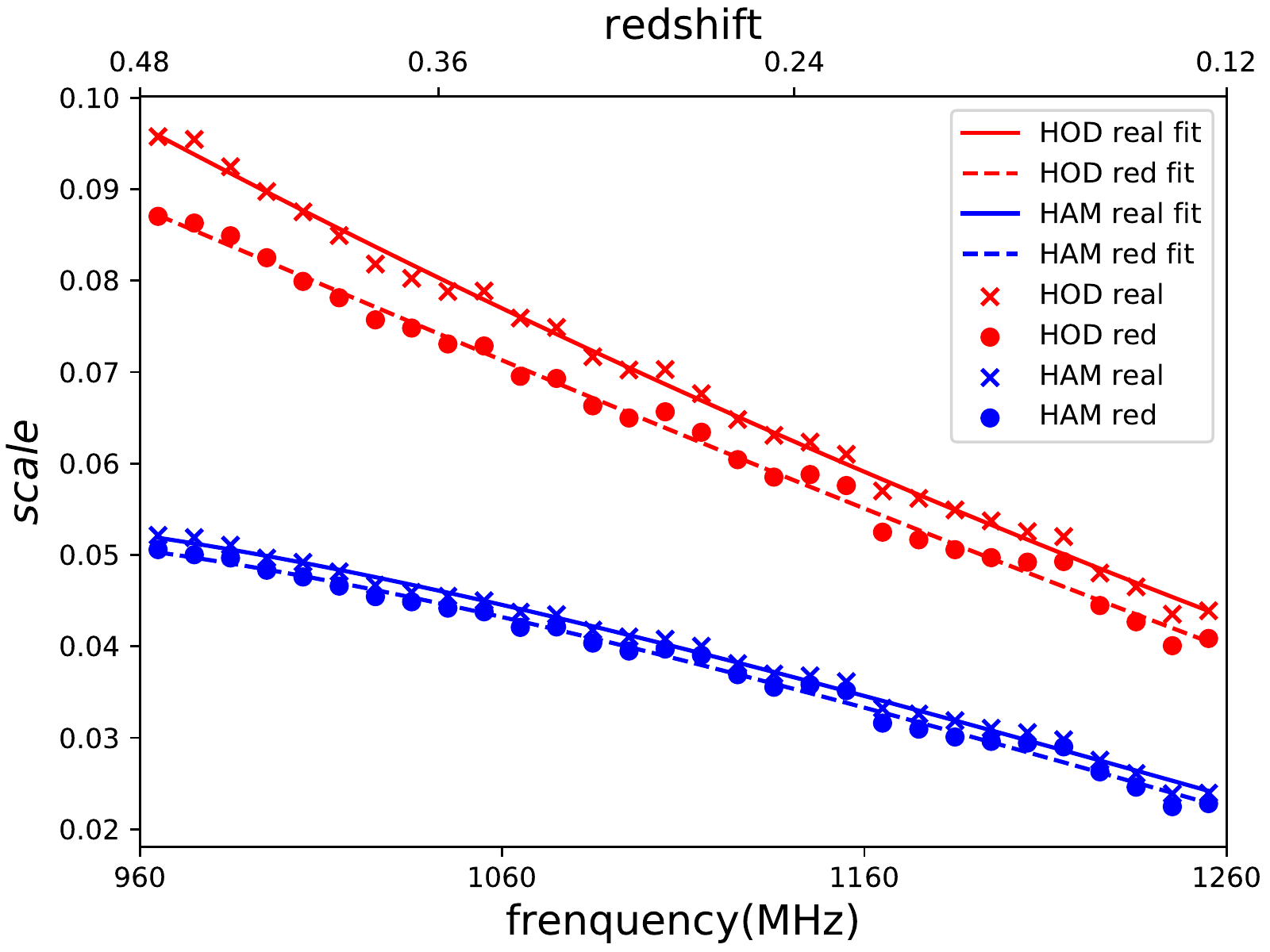}
\caption{In the upper left panel, we have shown the histogram of the pixels of the brightness temperature distribution of the redshift space mock map from 990 MHz to 1000 MHz smoothed by a 40 arcmin Gaussian kernel. After smoothing, the histogram can be very well fitted by the lognormal distribution. The fitted parameters, $s$, $loc$, and $scale$, are shown as a function of the redshift (frequency) in the other three panels. For each of them we have also fitted a parabola function, which turns out to be a good fit. The lognormal fitted parameters are shown in red color for the HOD mock and blue color for the HAM mock, crosses for the real space mock and dots for the redshift space mock. The parabola fitting curves are shown in red color for the HOD mock and blue color for the HAM mock, solid lines for the real space mock and dashed lines for the redshift space mock.}
\label{fig:lognorm}
\end{figure*}
The lognormal distribution can provide a reasonable fitting to the pixel histogram of the mock map, both in redshift space and in real space. Furthermore, after we considered the 40 arcmin Gaussian kernel smoothing for the mock map, we found that the lognormal function can fit the pixel histogram very well. We illustrated one slice as an example in Fig. \ref{fig:lognorm}. We fitted the histogram by a lognormal distribution using the stats routine in scipy \citep{scipy}. For every redshift bin, we fitted the histogram by lognormal distribution and showed the parameters in Fig. \ref{fig:lognorm}. The lognormal probability density function is
\begin{equation}
\begin{aligned}
    &f(y,s) = \dfrac{1}{sy\sqrt{2\pi}}\exp\left(-\dfrac{-\log^{2}(y)}{2s^{2}}\right) / \text{scale}\,, \\
    &y = (x - \text{loc}) / \text{scale}\,,
\end{aligned}
\end{equation}
where the $x$ variable is given by $T_b$ in our case.

We have used the parabola function, $f(z) = a + bz + cz^{2}$, to fit these three parameters of the lognormal distribution as a function of redshift $z$. The fitted results are given in Tab. \ref{tab:lognorm}.
\begin{table}[]
    \centering
    \begin{tabular}{|c|c|c|c|c|}
     Mock & parameter & $a$ & $b $& $c$  \\
     \hline
     HOD Real & s & 0.99 & -2.8 & 2.9 \\
     HOD Red  & s & 1.1  & -2.9 & 3.0 \\
     HAM Real & s & 1.6  & -4.4 & 4.6 \\
     HAM Red  & s & 1.7  & -4.7 & 4.9 \\
     HOD Real & loc & -0.0016  & -0.047 & -0.025 \\
     HOD Red  & loc & -0.00037 & -0.050 & 0.0046 \\
     HAM Real & loc & 0.00095  & -0.021 & 0.0030 \\
     HAM Red  & loc & 0.0015   & -0.024 & 0.0080 \\
     HOD Real & scale & 0.021  & 0.18   & -0.046 \\
     HOD Red  & scale & 0.017  & 0.19   & -0.081 \\
     HAM Real & scale & 0.0058 & 0.16   & -0.12  \\
     HAM Red  & scale & 0.0042 & 0.16   & -0.13  \\
    \end{tabular}
    \caption{The parabola fitting parameters of lognormal parameters as a function of redshift $z$}
    \label{tab:lognorm}
\end{table}

Given these parameters, one can generate a similar mock map as ours using lognormal distribution \citep{Xavier:2016}. It will be helpful for further study and comparison.
\subsection{Effect of Bin Size}\label{subsec:binsize}
\begin{figure}
    \centering
    \includegraphics[width=0.45\textwidth]{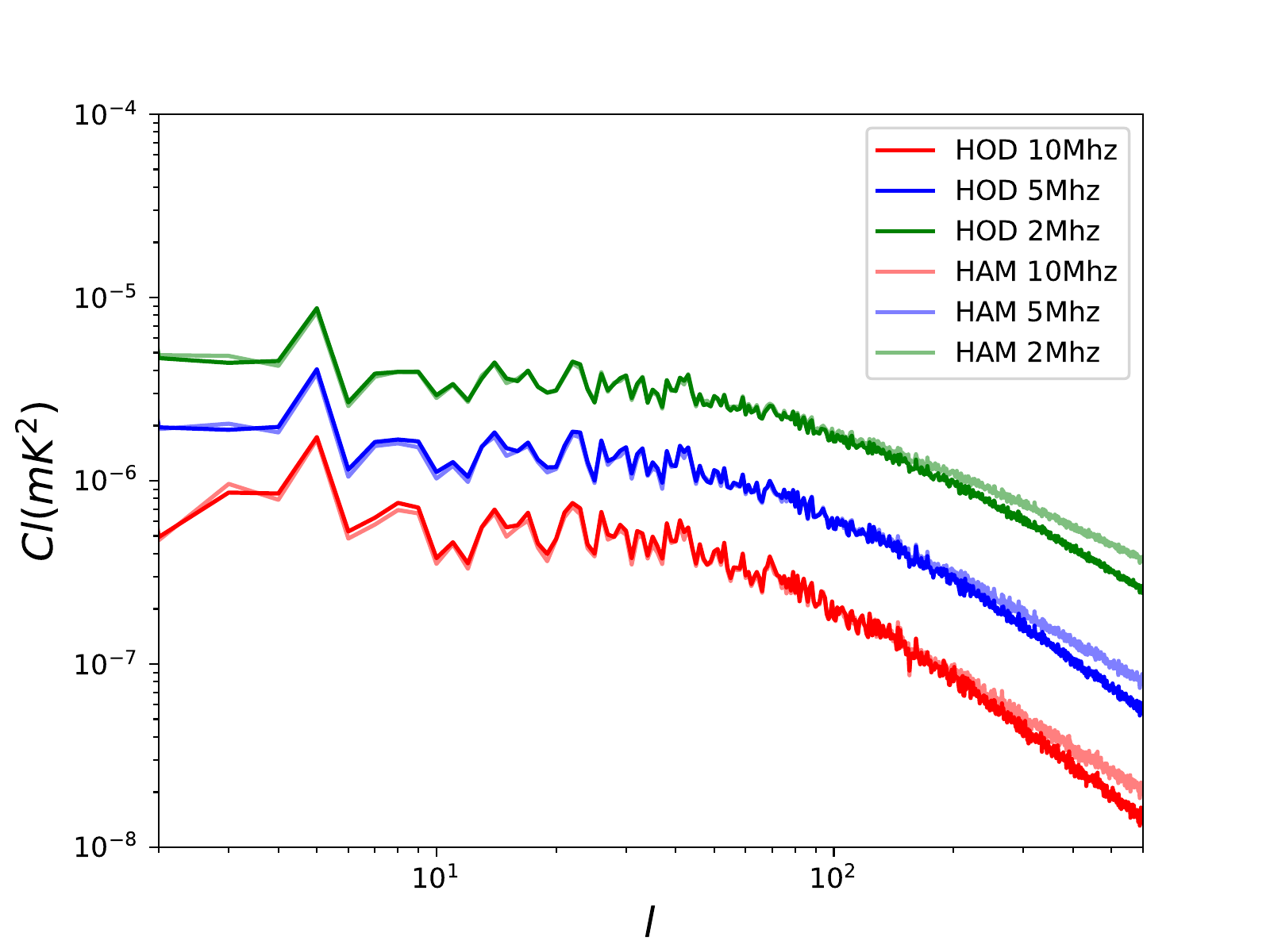}
    \includegraphics[width=0.45\textwidth]{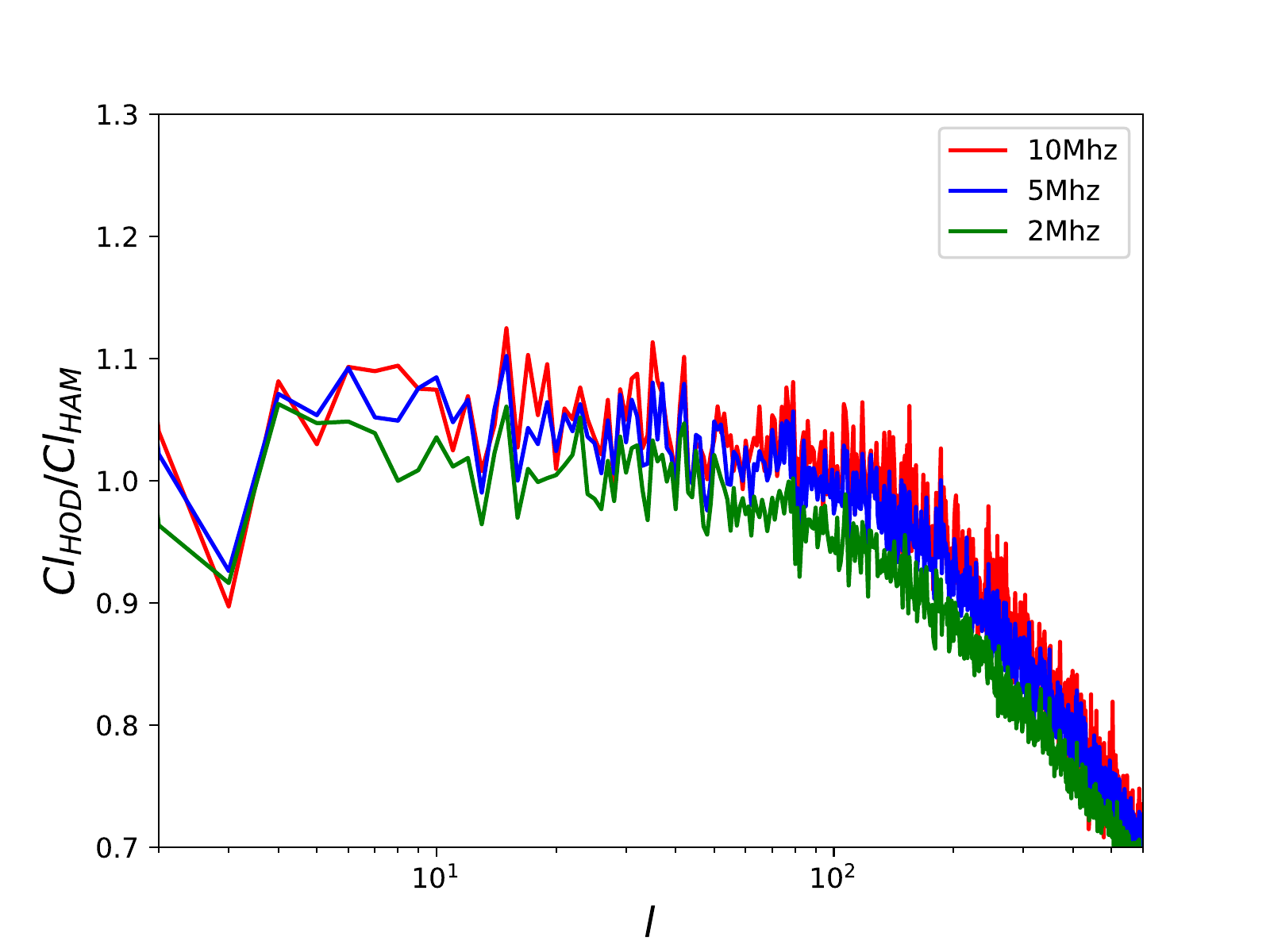}
    \caption{The comparison of HOD and HAM mocks in real space is shown with 10 (5, 2) MHz in red (blue, green) color. In the upper panel, the HOD results are shown in thick solid lines and the HAM results are shown in faint solid lines. Notice that for better illustration, we have artificially amplified the curves by 0.5 (1, 2) times for 10 (5, 2) MHz results. In the lower panel, we show the ratio of angular power spectrum between HOD mocks and HAM mocks. Clearly, HAM mocks have higher angular power spectrum at small scale and it depends weakly on the bin width.}
    \label{fig:binsize_hodham}
\end{figure}

As discussed in Sec. \ref{subsec:cls}, different ways of populating \hi\ gas in galaxies (halos) will lead to different bias and one-halo term, and the bin width of 10 MHz will largely wipe out the effect of FoG. One step further, it is interesting to ask the following question, if we use different bin width, for example 2 MHz or 5 MHz, will the conclusion be different? Can we see more significant effect of FoG or different small scale power introduced by one-halo term with smaller bin width? To answer this question, we have generated the mock map with HOD and HAM method in 2 MHz bin width and 5 MHz bin width, following the same steps as we used for generating the 10 MHz bin width mock map. 

In Fig. \ref{fig:binsize_hodham}, we show the comparison of angular power spectrum among the mocks with 10 MHz, 5 MHz and 2 MHz. For a more fair comparison, we take the frequency range from 990 MHz to 1000 MHz, which consists of 1 bin in the 10 MHz mock, 2 bins in the 5 MHz mock and 5 bins in the 2 MHz mock. We take the average of the angular power spectrum measured from these 1 (2, 5) bin(s) of 10 (5, 2) MHz mocks. We have also amplified the results to keep the curves of different bin width separated for better illustration. We can clearly see the higher angular power spectrum of HAM mocks compared to the HOD mocks at small scale ($l>100$). The angular power spectrum of HOD mocks comparing to HAM mocks are lower by more than $10\%$ at $l>300$. It is also obvious that the difference hardly depends on the bin width.  It is well expected, since no matter if the difference of small scale power is due to scale dependent bias or one-halo term, they are all isotropic. The width of frequency bin is expected to play little role in the power spectrum.
\begin{figure}
    \centering
    \includegraphics[width=0.45\textwidth]{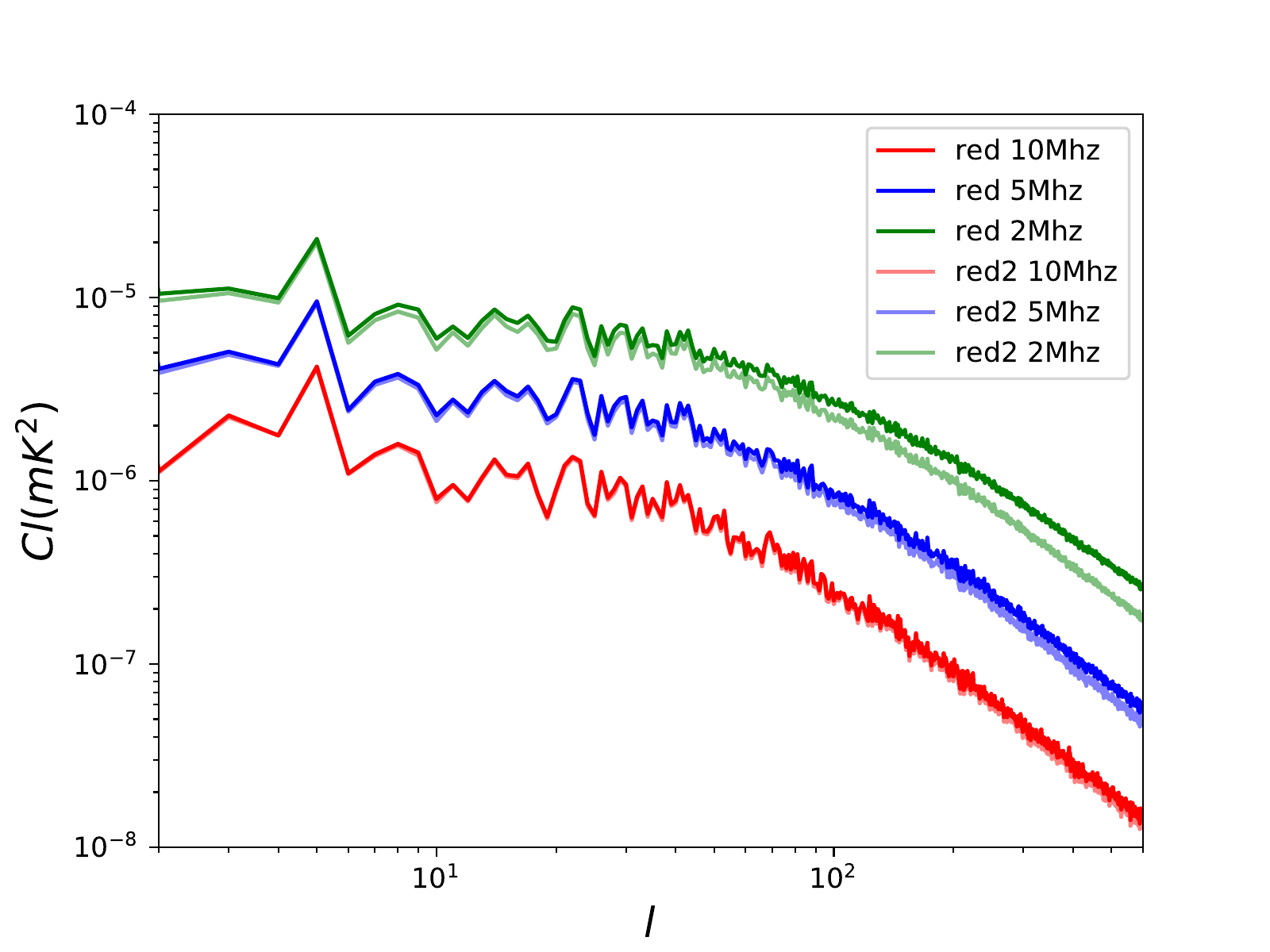}
    \caption{The comparison of "red" (point source assumption)  HOD mock and "red2" (velocity dispersion induced FoG effect modelled) HOD mock is shown with 10 (5, 2) MHz in red (blue,green) color. The "red" result are shown in thick solid lines and the "red2" results are shown in faint solid lines. Notice that for better illustration, we have artificially amplified the curves by 0.5 (1, 2) times for 10 (5, 2) MHz results. The FoG suppression at small scale is more clear with smaller bin width.}
    \label{fig:binsize_red}
\end{figure}

In contrast, the FoG effect is anisotropic. The FoG effect is pointing toward the line-of-sight direction. We expect to see stronger suppression due to FoG effect at small scale with smaller bin width, since there is less smoothing. The result is shown in Fig. \ref{fig:binsize_red}, which is exactly what we have expected. We have taken point source assumption in the "red" mock, and take the velocity dispersion into account in the "red2" mock. Therefore, the FoG effect is modelled in the "red2" sample. We can see that the FoG effect is very clear in the 2 MHz mock, weaker in the 5 MHz mock and almost indistinguishable in the 10 MHz mock. We can conclude that smaller frequency bin width is very effective in identifying the FoG effect in 21-cm intensity mapping observations. However, we should also notice that with smaller bin width, the noise level is also higher. Therefore, it is an open question whether we can identify the FoG effect in the real observations by BINGO.
\subsection{Comparison with {\tt UCLCl}}\label{subsec:uclcl}
\begin{figure}
    \centering
    \includegraphics[width=0.45\textwidth]{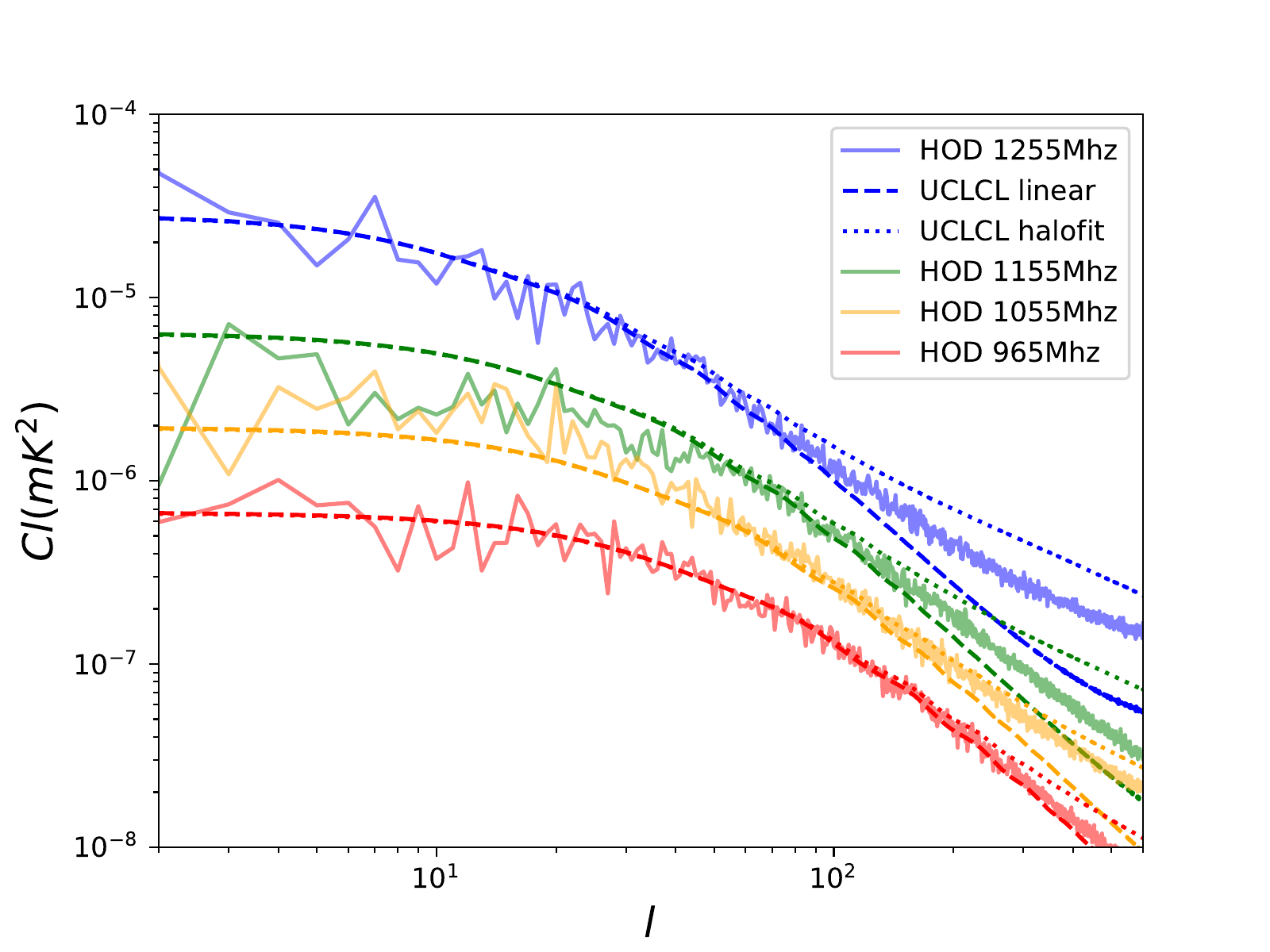}
    \caption{We show the angular power spectrum of HOD mocks (solid faint curves) comparing to the calculation of {\tt UCLCl} code with linear calculation (dashed lines) and halofit nonlinear correction (dotted lines) in redshift space. Red (orange, green, blue) lines show the result from 960 (1050, 1150, 1160) MHz to 970 (1060, 1160, 1260) MHz, middled at 965 (1055, 1155, 1255) MHz, which are amplified by 0.25 (0.5, 1, 2) times for better illustration. HOD mock measurements are higher than the linear calculation, but lower than the halofit calculation at small scale.}
    \label{fig:uclvshod}
\end{figure}

If we take the mock maps and the measured angular power spectrum as real observation, we need a fast model prediction to compare with the map and see whether we can fit it to the observation. It is essential for parameter fitting and cosmological parameter constraints.
{\tt UCLCl} (Unified  Cosmological  Library  for $C_\ell$'s)  is a package that calculates the angular power spectra \citep{uclcl1,uclcl2} from the power spectrum of density fluctuations through
\begin{equation}
    C_{l}^{ij}=4\pi \int \dfrac{dk}{k}W_{l}^{i}(k)W_{l}^{j}(k)\Delta^2(k)\,,
\end{equation}
where $W_{l}(k)$ is the window function that absorbs all the processes involved in the evolution, convolution and the projection effects, and $\Delta^2(k)$ is the dimensionless matter power spectrum. By using the linear perturbation calculation of the matter power spectrum, we can get the linear angular power spectrum. By using the halofit nonlinear correction \citep{takahashi2012halofit}, we can take the one-halo term nonlinear effect into account. We have calculated the angular power spectrum using the {\tt UCLCl} package with the cosmological parameters consistent with the HOD mock. 

In Fig. \ref{fig:uclvshod}, we have compared the results of the {\tt UCLCl} and our HOD mock measurements in redshift space. The point mass assumption was taken in this HOD mock for comparison. The HOD mock results are higher than the linearly calculated ones, but lower than the halofit corrected results at small scale. At large scale, the {\tt UCLCl} results are consistent with the HOD mock except for the obvious fluctuations due to cosmic variance. At lower redshift (higher frequency), we can see larger small scale nonlinear correction than at higher redshift (lower frequency). There are two reasons for that, one is that at higher redshift, the one-halo term plays a smaller role than at lower redshift, the other one is that at lower redshift, the shell of \hi\ distribution that we can measure is closer to us, which means that at lower redshifts we can see more details from small scales with the same angular resolution when compared to higher redshift shells. 

If we take the HOD mock as a benchmark, we can see that the linear case have underestimated the angular power spectrum at small scales and the halofit correction has overestimated the angular power spectrum. At higher redshift, the {\tt UCLCl} calculation is closer to the measurements of the HOD mock. At $l<80$, the {\tt UCLCl} linearly calculated results, which adopted the linear bias assumption, can fit to the HOD mock from 1250 MHz to 1260 MHz. From 960 MHz to 970 MHz, the {\tt UCLCl} linear calculation can fit the HOD mock result up to $l<200$. This comparison clearly tells us the shortage of our modelling for the 21-cm intensity mapping angular power spectrum using {\tt UCLCl} package. Without further improvement in the modelling at small scale and better understanding about the \hi\ distribution in halos, it is hard to use the full information of angular power spectrum at $l>80$ at the lowest redshift and $l>200$ at the highest redshift for BINGO.
\subsection{\hi\ Bias and Abundance}\label{subsec:bias}
\begin{figure}
    \centering
    \includegraphics[width=0.45\textwidth]{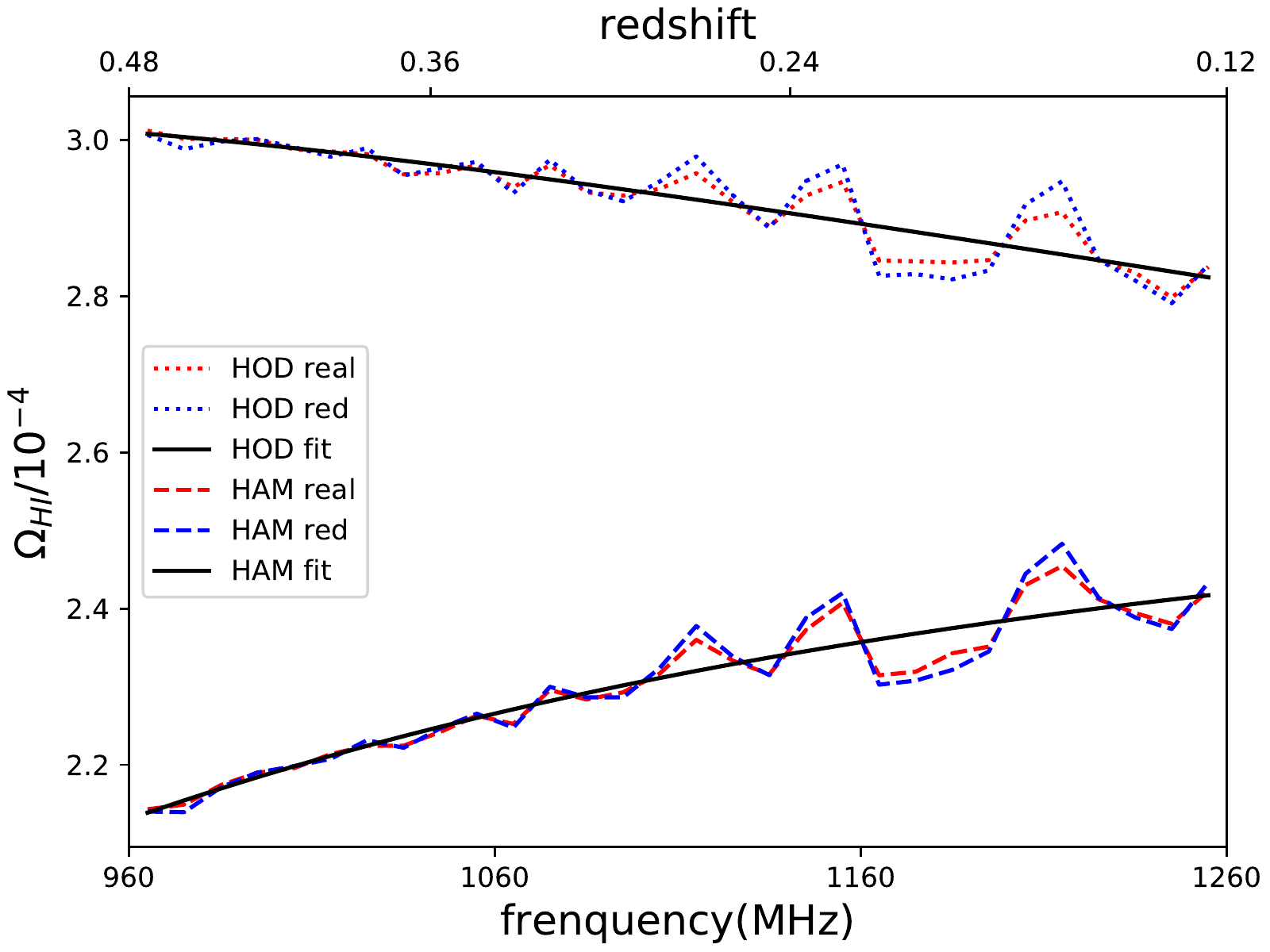}
    \caption{We show the $\Omega_{\rm \hi\,}$ as a function of frequency (redshift) in our mock map, the dotted lines show the result of the HOD mock and the dashed curves show the result of HAM mock. "real" means the real space mock, "red" means the redshift space mock under point source assumption. The black solid lines show the parabola fitting results to the real space mock.}
    \label{fig:omegahi}
\end{figure}
\begin{figure}
    \centering
    \includegraphics[width=0.45\textwidth]{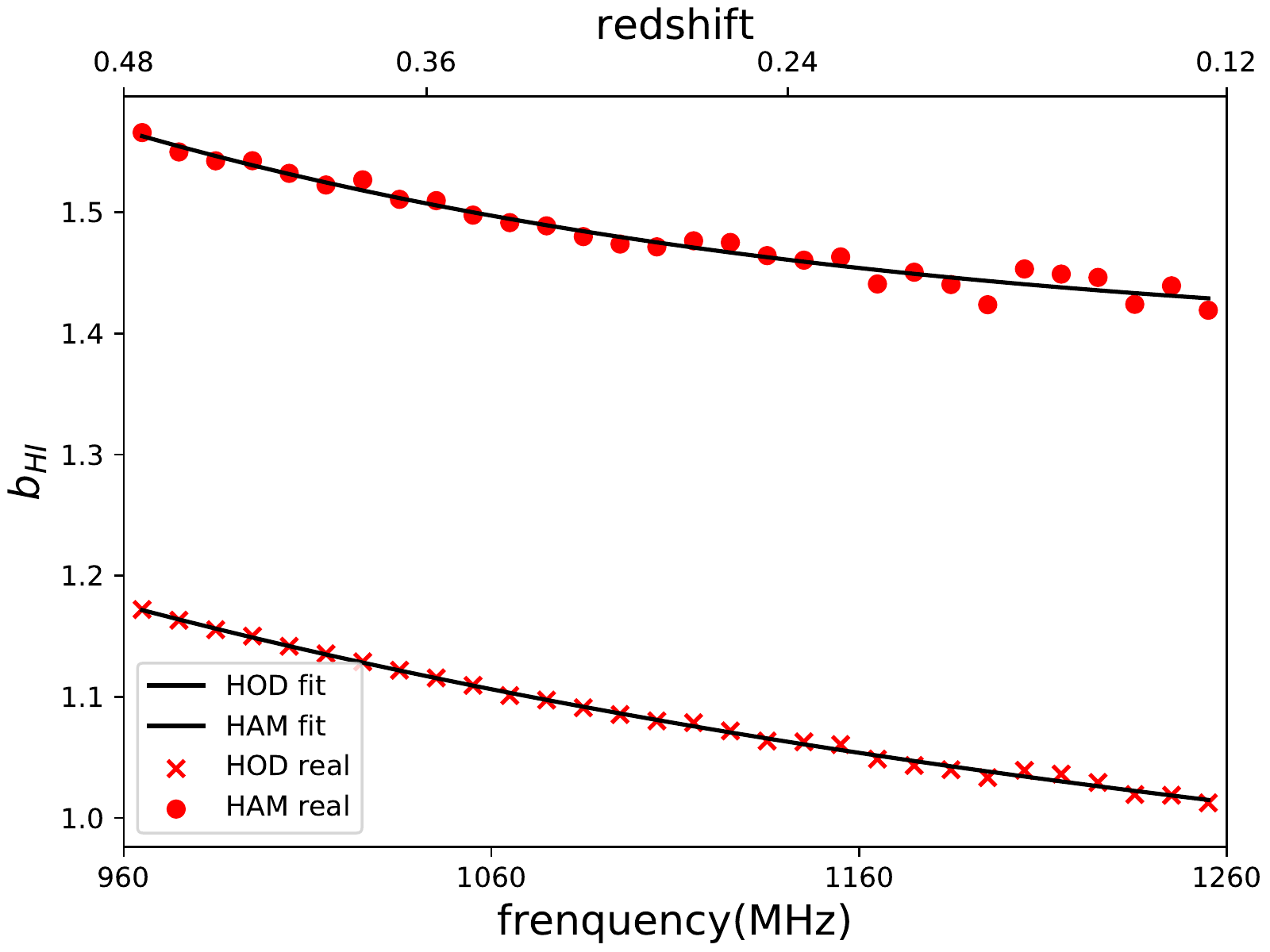}
    \caption{We show the $b_{\rm \hi\,}$ as a function of frequency (redshift) in our mock map, the crosses show the result of the HOD mock and the dots show the result of the HAM mock. Both the real space and redshift space mock bias are measured, and they are almost identical. The solid black lines show the parabola fitting results.}
    \label{fig:bias}
\end{figure}

The overall amplitude of the angular power spectra of 21-cm intensity mapping is closely related to $\Omega_{\rm \hi\,}$ and $b_{\rm \hi\,}$. Therefore,  observations can give constraints on these parameters. The detailed forecast is provided in our companion paper VII \citep{2020_forecast}. Here we provide the "answer" of the mock map for the future parameter fitting and pipeline building. We calculate the $\Omega_{\rm \hi\,}$ in each mock map by adding up all the \hi\ mass in each redshift bin, and then divided by the volume of each bin. We show $\Omega_{\rm \hi\,}$ as a function of frequency (redshift) in Fig. \ref{fig:omegahi}. The measurement of HOD mock is shown in dotted lines and the HAM mock is shown in dashed lines. The population of \hi\ mass is different between the real space mock map and the redshift space mock map in the same frequency bin due to RSD effect. We have illustrated the result of real space mock and redshift space mock in different colors, but clearly there is little difference between the real space and redshift space. The RSD has little impact in the $\Omega_{\rm \hi\,}$ in each bin. We have provided a parabola fitting to the real space mocks, which gives
\begin{equation}
    (\text{HOD}) \Omega_{\rm \hi\,} = 2.7\times 10^{-4} + 1.0\times 10^{-4}z - 8.0\times 10^{-5}z^{2},
\end{equation}
for the HOD mock and
\begin{equation}
    (\text{HAM}) \Omega_{\rm \hi\,} = 2.5\times 10^{-4} - 4.0\times 10^{-5}z - 7.0\times 10^{-5}z^{2},
\end{equation}
for the HAM mock, shown in the black solid lines.

For the HOD mock, we use the publicly available python package COLOSSUS \citep{colossus} to calculate the bias of each halo using the empirical relation provided by \citet{tinker2010bias}, and by considering the following relation in real space
\begin{equation}
    \dfrac{Cl_{\rm HAM}}{Cl_{\rm HOD}} = \dfrac{\bar{T}_{\rm HAM}^2}{\bar{T}_{HOD}^2}\dfrac{b_{\rm HAM}^2}{b_{\rm HOD}^2}\; ,
\end{equation}
we estimated the bias of the HAM mock. We show the bias $b_{\rm \hi\,}$ as a function of frequency (redshift) in Fig.  \ref{fig:bias}. The crosses represent the result of the HOD mock and the dots represent the result of the HAM mock. Since the real space and redshift space results are almost identical, we just show the results of real space mock for better illustration. Similarly, we have also provided the parabola fitting results 
\begin{equation}
    (\text{HOD}) b_{\rm \hi\,} = 0.96 + 0.36z + 0.17z^{2},
\end{equation}
 for HOD mock and
\begin{equation}
    (\text{HAM}) b_{\rm \hi\,} = 1.4 + 0.092z + 0.50z^{2},
\end{equation}
for HAM mock, shown in the black solid lines. The big difference among the bias of the HOD mock and the HAM mock shows that different ways of \hi\ gas occupation can lead to very different linear bias, which is observable in the angular power spectrum. The 21-cm intensity mapping can also tell us about the \hi\ gas distribution, related to the galaxy formation and evolution.

There are three advantages of using the fitting formula:
\begin{itemize}
    \item it is useful for reproducing the map with other methods, such as using Gaussian realization or FLASK, which can easily generate large amount of realizations for constructing covariance matrix;
    \item it can largely reduce the number of free parameters about $\Omega_{\rm \hi\,}$ and $b_{\rm \hi\,}$ with fitting formula, when using maps at many different redshifts;
    \item this fitting formula is independent of the number of bins, it is useful when we need to change the number of bins. 
\end{itemize}
\section{Summary and Discussion}\label{sec:summary}
In this paper, we have introduced the method of building mock 21-cm intensity maps using N-body simulation data. In summary, we have achieved the following points,
\begin{itemize}
    \item[1] We have used the ELUCID N-body simulation and its semi-analytical galaxy catalog to study the \hi\ halo occupation distribution. We have concluded that for halos less massive than $10^{12}M_{\odot}/h$, the major contribution of \hi\ gas is from the central galaxies. While for halos more massive than $10^{13}M_{\odot}/h$, the major contribution of \hi\ gas is from the satellite galaxies. Neglecting the \hi\ contribution in satellite galaxies is not a good approximation.
    \item[2] We have provided a good fitting formula of \hi\ mass-halo mass relation, make it easier to populate \hi\ gas into dark matter halos. The fitting function is given by Eq. (\ref{eq:hihalofit}).
    \item[3] The HOD method was verified using the ELUCID simulation by comparing to the SAM galaxy catalog, within $10\%$ accuracy, the power spectrum of the HOD repopulated \hi\ distribution is correct up to $k=0.4h\text{Mpc}^{-1}$.
    \item[4] The mass resolution of the N-body simulation can lead to scale dependent bias. With the halo mass limit down to around $10^{11}M_{\odot}/h$, the linear bias assumption, which assume a constant bias number, is still valid. The linear bias assumption is not valid with $10^{12}M_{\odot}/h$ halo mass limit. This sets a resolution requirement for N-body simulation to be used to build \hi\ mock map.
    \item[5] Using HR4 simulation, we have built the full-sky 21-cm intensity mock maps covering from 960 MHz to 1260 MHz, covering the range of BINGO (from 980 MHz to 1260 Mhz). Both the HOD method and abundance matching (HAM) method are applied to get two kinds of mocks.
    \item[6] The redshift space distortion (RSD) was taken into consideration, we introduced the method to consider both the point source redshift space distortion and the method to consider the velocity dispersion to account for finger-of-god (FoG) effect.
    \item[7] The lognormal distribution can provide a reasonable fitting to the pixel histogram of the mock map, if the map is smoothed by 40 arcmin Gaussian kernel, the fitting will be very good. 
    \item[8] We have provided the lognormal fitting parameters for our mock maps. A further parabola fitting to the lognormal parameters as functions of redshift was done. The fitting results are given. It is easy to realize a similar brightness temperature distribution with lognormal distribution using our parameters. 
    \item[9] We have discussed about the effect of Gaussian beam smoothing, RSD and the difference of HOD mock and HAM mock by comparing the angular power spectrum. Different bias and Kaiser effect can be discriminated. The nonlinear effect at small scale and FoG effect will not be important if the 40 arcmin Gaussian smoothing is taken into account.
    \item[10] The effect of different frequency bin sizes was discussed. FoG effect can be more easily identified with smaller bin size. Different bin size has little impact in telling the difference between HOD mock and HAM mock.
    \item[11] The theoretical calculation was done using {\tt UCLCl} package and compared to our mock map measurements. We found that the {\tt UCLCl} results can fit to the mock at large scale, but failed at small scale. The lower the redshift, the worse. They fit well at $l<80$ for the highest frequency bin and at $l<200$ for the lowest frequency bin of BINGO. The linear theory calculation has underestimated the angular power spectra at small scale and the halofit nonlinear correction has overestimated it. We need a better modelling to make full use of the whole range of angular power spectra.
    \item[12] We have provided the measurement of $\Omega_{\rm \hi\,}$ and $b_{\rm \hi\,}$ as a function of frequency (redshift) in the HOD mock and HAM mock. The parabola function can provide very good fitting to the measurements. The fitted functions in each case are listed in Sec.  \ref{subsec:bias}.
\end{itemize}

Beside constraining cosmological parameters, 21-cm intensity mapping is expected to constrain the astrophysical processes related to \hi\ gas. There is no conclusion about the best method, HAM and HOD, two different methods applied in this paper for populating \hi\ gas in haloes, can be regarded as valid tests. Since we can distinguish HAM from HOD using 21-cm intensity mapping by constraining the bias parameter, it proves the ability of discriminating the \hi\ population in haloes. The mock we have generated here will serve as the source signal to test the BINGO data analysis pipeline. This will include parameter fitting, BAO signal fitting and many other methods developed for intensity mapping.

The methodology we have introduced here will be useful in building mocks based on fast simulations such as {\tt COLA-HALO} \citep{koda2016fast}. After generating hundreds of mocks, we can measure the covariance matrix, which is essential for estimating the cosmological parameters. On the other hand, this methodology can also be applied for other 21-cm intensity mapping experiments such as SKA. Next step, we will combine the foreground, noise and mask of BINGO to generate a more realistic mock map for testing our data analysis pipeline. In the future, we will also include more effects that might change the map, such as point radio sources, lensing effects, gravitational redshift etc. By making more and more realistic mock maps, we can improve the understanding about intensity mapping and provide better data challenge for the pipeline.
\begin{acknowledgements}
      This work was supported by the Ministry of Science and Technology of China (grant nos. 2018YFA0404601) and National Natural Science Foundation of China (grant nos. 11973033, 11835009). 
      The BINGO project is supported by FAPESP grant 2014/07885-0; the support from CNPq is also gratefully acknowledged (E.A.). 
      C.A.W. acknowledges a CNPq grant 2014/313.597. 
      T.V. acknowledges CNPq Grant 308876/2014-8. 
      J.Z. was supported by IBS under the project code, IBS-R018-D1. 
      C.P.N. would like to thank S{\~a}o Paulo Research Foundation (FAPESP), grant 2019/06040-0, for financial support. 
      A.A.C. acknowledges financial support from the China Postdoctoral Science Foundation, grant number 2020M671611. 
      F.B.A. acknowledges the UKRI-FAPESP grant 2019/05687-0, and FAPESP and USP for Visiting Professor Fellowships where this work has been developed.
      R.G.L. thanks  CAPES (process 88881.162206/2017-01) and the Alexander von Humboldt Foundation for the financial support. 
      A.R.Q., F.A.B., L.B., and M.V.S. acknowledge PRONEX/CNPq/FAPESQ-PB (Grant no. 165/2018).
      H.X. and Z.Z. are supported by the Ministry of Science and Technology of China (grant No. 2018YFA0404601 and 2020SKA0110200), and the National Science Foundation of China (grant Nos. 11621303, 11835009 and 11973033).
      K.S.F.F. thanks S{\~a}o Paulo Research Foundation (FAPESP) for financial support through grant 2017/21570-0.
      V.L. acknowledges the postdoctoral FAPESP grant 2018/02026-0.
      L.S. is supported by the National Key R\&D Program of China (2020YFC2201600).
\end{acknowledgements}

%
%
\bibliographystyle{aa}
\bibliography{himock}

\end{document}